\documentclass[]{aa} % for the letters  

\usepackage{graphicx}
\usepackage{txfonts}
\usepackage{natbib}
\usepackage{longtable}
\usepackage{lscape}
\usepackage{color}

\providecommand{\sorthelp}[1]{}

\newcommand{\Planck}{{\it Planck }}

\newcommand{\nL}{$n_{\rm L}$}
\newcommand{\nP}{$n_{\rm P}$}
\newcommand{\nI}{$n_{\rm I}$}

\begin{document}
 
\title{
%Radiative transfer modelling of \Planck cold clumps with SOC radiative
%transfer program
SOC program for dust continuum radiative transfer
}

\author{Mika  Juvela}

\institute{
Department of Physics, P.O.Box 64, FI-00014, University of Helsinki,
Finland, {\em mika.juvela@helsinki.fi}
}

\authorrunning{M. Juvela et al.}

\date{Received September 15, 1996; accepted March 16, 1997}

\abstract { 
%% BACKGROUND
Thermal dust emission carries information on physical conditions and
dust properties in many astronomical sources. Because observations
represent a sum of emission along the line of sight, their
interpretation often requires radiative transfer modelling.
} 
{
% AIMS
We describe a new radiative transfer program SOC for computations of
dust emission and examine its performance in simulations of
interstellar clouds with external and internal heating.
}
% METHODS
{
SOC implements the Monte Carlo radiative transfer method as a parallel
program for shared memory computers. It can be used to study dust
extinction, scattering, and emission. We tested SOC with realistic
cloud models and examined the convergence and noise of the dust
temperature estimates and of the resulting surface brightness maps.
}
% RESULTS
{
SOC has been demonstrated to produces accurate estimates for dust
scattering and for thermal dust emission. It performs well with both
with CPUs and with GPUs, the latter providing up to an order of
magnitude speed-up. In the test cases, ALI improved the convergence
rates but also was sensitive to Monte Carlo noise. Run-time refinement
of the hierarchical-grid models did not help in reducing the run times
required for a given accuracy of solution. The use of a reference
field, without ALI, works more robustly. It also allows the run time
to be optimised if the number of photon packages is increased only as
the iterations progress.
}
%% CONCLUSIONS
{ 
The use of GPUs in radiative transfer computations should be
investigated further.
}

\keywords{
Radiative transfer -- ISM: clouds -- Submillimetre: ISM -- dust, extinction -- 
Stars: formation
}

\maketitle

\section{Introduction} \label{sect:intro}

Most knowledge of astronomical sources is based on radiation that is
produced and processed by inhomogeneous objects viewed from a single
viewpoint. This also applies to interstellar medium, the line emission
from gas and the continuum emission from interstellar dust. Radiative
transfer (RT) defines the relationships between the physical
conditions in a radiation source and the observable radiation. Thus,
RT modelling is needed to determine the source properties or, more
generally, the range of properties consistent with observations.

Many RT programs exists for the modelling of dust continuum data
\citep[e.g.][]{Bianchi1996, Dullemond2000, Gordon2001,
BjorkmanWood2001, Juvela2003, Steinacker2003, Wolf2003, Harries2004,
Ercolano2005, Juvela2005, Pinte2006, Jonsson2006, Chakrabarti2009,
Robitaille2011, Lunttila2012, Natale2014, Baes2015_SKIRT} and
benchmarking projects have quantified the consistency between the
different methods and implementations \citep{Ivezic1997, Pascucci2004,
Gordon2017_TRUST-I}. Part of these codes can also be used in studies
of polarised scattered or emitted radiation \citep{Whitney2003,
Pelkonen2007, Bethell2007, Reissl2016, Peest2017}.

RT is computationally demanding because each source location is in
principle coupled with all the other positions
\citep{Steinacker2013_ARAA}, the coupling changing with wavelength.
Emission calculations may need iterations where temperature estimates
and estimates of the radiation field are updated alternatingly. Most
codes for dust emission and scattering modelling are based on the
Monte Carlo method, the simulation of large numbers of photon packages
that represent the actual radiation field. In ``immediate
re-emission'' codes \citep{BjorkmanWood2001}, every interaction with
the medium leads to a re-evaluation of the dust emission from the
corresponding model cell. In other Monte Carlo codes the information
about the absorbed energy is stored during the simulation step, after
which the temperatures of all cells are updated. Models with high dust
temperatures and high optical depths may require a significant number
of iterations.

In the case of large 3D models, the run times become long and some
parallelisation of the computations may be necessary
\citep{Robitaille2011,Verstocken2017}. The Monte Carlo method allows
straightforward parallelisation of the radiation field simulations, at
the level of individual photon packages or between frequencies (if the
basic simulation scheme this allows) or even based on the
decomposition of the computational domain \citep{Harries2015}.
However, naive parallelisation -- where each computing unit performs
independent simulations with different random numbers -- may also be
the most efficient.

The parallelisation between nodes is typically handled with Message
Passing Interface (MPI\footnote{http://mpi-forum.org/})
and within a node (on a single shared-memory computer), for
example, with OpenMP\footnote{www.openmp.org}, lower level threads, or
other language-specific (or vendor-specific) tools.
Further speed-up could be obtained with graphics processing units
(GPUs) that theoretically are capable of more floating point
operations per second than even high-end CPUs. The use of GPUs has
been hindered by the more complex programming model and the limited
device memory. However, the massive parallelism provided by GPUs is
well suited for radiative transfer calculations and some applications
to astronomical radiative transfer already exist \citep{Heymann2012,
Malik2017}. These use the CUDA parallel programming platform, which is
proprietary to NVIDIA and cannot be used with GPUs of other
manufacturers. Furthermore, CUDA programs are written explicitly for
GPUs and cannot be run on CPUs.  GPU programming is becoming more
approachable as directive-based programming models are becoming
available. The support for GPUs (and ``accelerators'' such as
Intel Xeon Phi, AMD Radeon Instinct, etc.) is maturing in OpenMP. This
enables heterogeneous computing, the same program running on both GPUs
and CPUs. However, in this paper we describe the continuum radiative
transfer program SOC that is written using OpenCL
libraries\footnote{https://www.khronos.org/opencl/}. OpenCL is an open
standard for heterogeneous computing. Thus, SOC can in principle be
run unaltered on CPUs, GPUs, and even other accelerators. OpenCL is
supported by many vendors and, furthermore, has fully open source
implementations, making it more future-proof against the current rapid
changes and evolution of the GPU computing frameworks.

SOC has been compared to other continuum radiative transfer programs
in \citet{Gordon2017_TRUST-I}. In this paper we describe in more
detail some of the implementation details and examine the efficiency
of SOC and some of the implemented methods in the modelling of dust
emission from interstellar clouds clouds.  In a future paper,
SOC will be used to produce synthetic surface brightness maps for a
series of MHD model clouds and, based on these, to construct synthetic
source catalogues with the pipeline used for the original \Planck
Galactic Cold Cores Catalogue (PGCC) \citep{PGCC}. One of these MHD
models is already used in the tests in the present paper.

The contents of the paper are the following. In
Sect.~\ref{sect:methods} we discuss the implementation of the SOC
programme. In the results section Sect.~\ref{sect:results}, we
describe tests on the SOC performance. These include in particular
problems that require iterations to reach final dust temperature
estimates (Sect.~\ref{sect:iterations}-\ref{sect:accel}).  The
findings are discussed in Sect.~\ref{sect:discussion} and the final
conclusions are listed in Sect.~\ref{sect:conclusions}. In
Appendix~\ref{sect:SHG} we discuss further the calculation of emission
from stochastically heated grains.

\section{Methods} \label{sect:methods}

In this section, we present some details of the SOC implementation and
the model clouds used in the subsequent tests.

\subsection{SOC radiative transfer program} \label{methods:SOC}

SOC is a Monte Carlo radiative transfer program for the calculation of
dust emission and scattering. It has been used in some publications
\citep{Gordon2017_TRUST-I, Juvela2018_clumps, Juvela2018_G35} but we
describe here in more detail some of the implementation details.

\subsubsection{Basic program}

SOC uses OpenCL libraries, enabling the program to be run on both CPUs
and GPUs. This has affected some of the design decisions. The program
is run on a $host$, which calls specific routines, called kernels,
that are executed on a $device$. The host is the normal computer (CPU)
while the device can be a CPU (using the same resources as the host),
a GPU, or other accelerator. Thus, the memory available on the device
may be more limited than usual. SOC has separate kernels to carry out
the simulation of photon packages at a single frequency, to solve the
dust temperatures based on the computed absorbed energy
(non-stochastically heated grains only), and, based on that solution,
to produce surface brightness maps at given frequencies.

In \citet{Gordon2017_TRUST-I} we used an earlier version that employed
modified Cartesian grids. Current SOC uses cloud models defined on
hierarchical grids. The root grid is a regular Cartesian grid with
cubic cells. Refinement is based on octrees where each cell can be
recursively divided into eight sub-cells of equal size. In the
following we refer to such a set of eight cells as octet and the
number of hierarchy levels as \nL. The root grid is the level $L=1$
and a grid consisting only of the root grid has \nL=1. The cells of
the model cloud form a vector that starts with the cells of the root
grid, followed by all cells of the first hierarchy level, an so forth.
The links from parent cells to the first cell in the sub-octet are
stored in the same structure, encoding the index of the first child
cell as a negative value in place of the density value in the parent
cell. Because each octet is stored as consecutive elements of the
density vector, smaller auxiliary arrays are sufficient to store the
reverse links from the first cell of an octet to its parent cell. 
Neighbouring cells could be located faster by using explicit
neighbour lists \citep{Saftly2013}. However, to reduce memory
requirements, an important consideration especially on GPUs, we are
currently not using that technique.
Therefore, each time a photon package is moved to a new cell, a
partial traversal of the hierarchy is required, up to a common parent
cell and then down to the new leaf node that corresponds to the next
cell along the path of the photon package.

The SOC program is based on the usual Monte Carlo simulation where the
radiation field is simulated using a pre-determined number of photon
packages, each standing for a large number of real photons. SOC
concentrates on the radiative transfer problem and does not have
built-in descriptions of specific dust models, radiation sources, or
cloud models (density distributions). These inputs are read from files
that are specified in the SOC initialisation file. They include the
cloud hierarchy with the density values, the dust cross sections for
absorption and scattering, and scattering functions tabulated as
functions of frequency and scattering angle.

The simulation is done using a fixed frequency grid. Therefore, at
each step of the calculations, the kernel only needs data related to a
single frequency. These include the densities and, for the current
frequency, the optical depth and a counter for the number of absorbed
photons. The density vector (one floating point number per cell)
includes basic information about the grid geometry. The information
about parent cells is less than one eight of this (link for first cell
of each octet, excluding the root level). The host sets up the data
for the current frequency and these are transferred to the device.
After the kernel has completed the simulation of the radiation field,
the information of the number of absorptions in each cell can be
returned to the host. The host loops over simulated frequencies, calls
the kernel to do the simulations, and gathers information of
absorption events. 

For stochastically heated grains, the information about absorbed
energy is converted to dust emission using an external programs such
as DustEM \citep{Compiegne2011} or the program we used in
\citet{Lunttila2012} and in \citet{Camps2015}. On the host side this
requires the storage of large arrays that, for modern computers, would
not set serious limits to the size of the computed models. However, in
SOC these are stored as memory-mapped files. 
Memory-mapped files reside on disk but can be used in the program
transparently as normal arrays. The operating system is responsible
for reading and writing the data, as needed, without the data
ever being in the main memory all at the same time. 
Thus the main limitation remains the device memory and the amount of
data per a single frequency. 

If grains are assumed to be in an equilibrium with the radiation
field, the emission is calculated by SOC itself, again using a
separate kernel. The integral of the absorbed energy over frequency
can be gathered directly during the simulations, without the need for
large multi-frequency arrays. The absorption information remains on
the device, and only the computed dust emission is returned to the
host. As the final step, the host calls a mapping kernel to make
surface brightness maps for the desired frequencies and directions.

SOC can also produce images of scattered light. However, these
are usually made with separate runs that employ methods such as forced
first scattering, peel-off, and potentially additional weighting
schemes \citep{Juvela2005, Baes2016_biasing}. Apart from the use of
different spatial discretisation, SOC results for scattering problems
have already been discussed in \cite{Gordon2017_TRUST-I}.  We
concentrate in this paper on the dust emission and especially the
methods described in the next section.

\subsubsection{Additional methods}

SOC implements some features beyond the basic Monte Carlo Scheme.
These include forced first scattering and accelerated lambda
iterations (ALI). ALI (also called accelerated Monte Carlo or AMC) is
computed using a diagonal operator that solves explicitly for the
cycle of photons that are absorbed within the emitting cell
\citep[see][]{Juvela2005}. The use of ALI incurs an additional cost of
storing on the device one additional floating point number per cell,
which is needed to keep track of the photon escape probabilities for
each cell. SOC includes optional weighting for the scattering
direction \citep{Juvela2005} and the step length between scattering
events. The latter is similar in nature to the method described in
\citep{Baes2016_biasing} but, to avoid explicit integration to the
model boundary, the probability distribution of the simulated free
path is based on the sum of two exponential functions of optical
depth, with user-selected parameters. The weighting is applied only on
the first step of each photon package. 

SOC can use a reference field to reduce the noise, especially in
calculations that involve several iterations \citep{Bernes1979,
Juvela2005}. In this method, each simulation step only estimates a
correction to the average radiation field determined by the previous
iterations. This requires that the absorptions caused by the reference
field are also stored, adding storage requirements by one number per
each cell and frequency. For grains at an equilibrium temperature,
this reduces to a single number per cell (the total absorbed energy).
In an ideal case, the overall noise would then decrease as
\nI$^{-1/2}$, as the function of the number of iterations -- assuming
that the number of photon packages per iteration, \nP, is constant.
However, because also the temperatures change during the iterations,
the noise is likely to decrease more slowly, especially during the
first iterations.

Iterations are needed if the model includes high optical depths and
dust is heated to such high temperatures that its emission no longer
freely escapes the model volume. In the context of interstellar
clouds, this means dust near embedded radiation sources, possibly in a
very small fraction of the whole model volume. In these cases, the
reference field can be used for further optimisation. After the first
iteration, the reference field already contains the information of all
constant sources, such as the embedded point sources or the background
radiation, and they can be omitted from the subsequent iterations.
Furthermore, it is possible to divide the cells to two categories.
Passive cells are those whose emission (resulting from their
temperature change during the iterations) is too low to affect the
temperature of other cells. Active cells are correspondingly those
whose temperature does change with iterations, affecting the
temperature of other cells. Once the former are included in the
reference field (if their emission indeed is significant at all), on
subsequent iterations only emission from the active cells needs to be
simulated. In SOC, the host can examine the changes of emission
between iterations and can thus omit the creation of photon packages
for cells for which the emission has not changed. However, in practice
the same is accomplished by weighted sampling where the number of
photon packages emitted from each cell is determined by the cell
luminosity. 

Finally, SOC has the option to change the spatial resolution also
during iterations. One could start with a low-resolution model, doing
fast iterations, and only later refine to the final resolution needed.
This could result in savings in the run times, depending on the
optical depths and the intensity of the dust re-emission. On the other
hand, a lower spatial resolution means optical depths for individual
cells, which tends to slow the convergence of temperature values down,
especially when ALI is not used.
%%%
The run-time refinement is briefly tested in Sect.~\ref{sect:accel}

The present SOC does not carry out the radiative transfer simulation
with polarised radiation but can nevertheless be used to calculate
synthetic maps of polarised dust emission. SOC can determine for each
cell the intensity and the anisotropy of the radiation field. Other
scripts are used to calculate the resulting polarisation reduction
factors, for example, according to the radiative torques theory, thus
also including the information about the magnetic field orientation.
The procedure is essentially identical to the computations presented
in \citet{Pelkonen2009}. The information of the polarisation reduction
and of the magnetic field geometry are read into SOC, which then
produces synthetic maps of the Stokes parameters. The calculations
ignore the effects on the total intensity that result from the cross
sections being dependent on the angle between the magnetic field and
the direction of the radiation propagation. This is usually not a
severe approximation (especially when compared to the many other
sources of uncertainty). However, based on SOC, another program is now
in preparation where all calculations are done using the full Stokes
vectors, taking into account the degree to which the grains are in
each cell aligned with the magnetic field \citep{Pelkonen2018}.

\subsection{Test model clouds}  \label{methods:model}

The first tests were performed with spherically symmetric density
distributions that were sampled onto hierarchical grids (see
Sect.~\ref{sect:1D}). 

\begin{figure}
%% \sidecaption
\includegraphics[width=8.8cm]{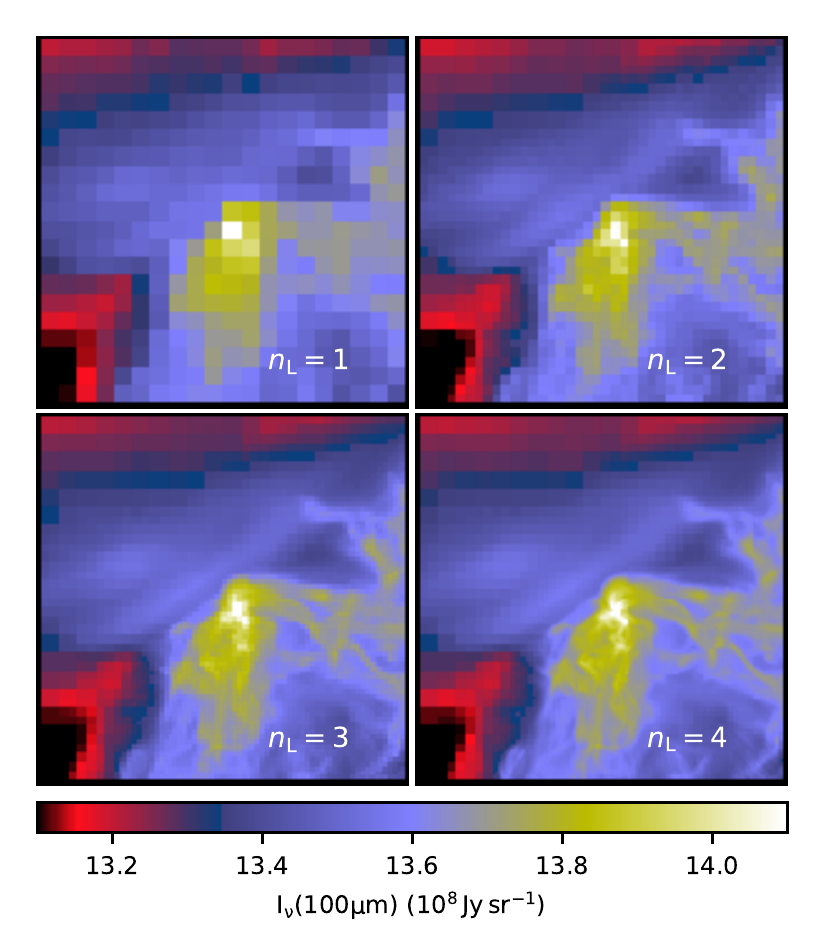}
\caption{
Examples of 100\,$\mu$m surface brightness maps computed for 3D model
clouds with a maximum refinement to \nL=1-4 hierarchy levels. In
calculations involving internal heating (as in these examples), point
source is placed in an empty root grid cell at the centre of the model
volume.
}
\label{fig:maps}
\end{figure}

Most tests employed a snapshots from the MHD simulations described in
\citet{Padoan2016_SN-I}, which has already been used for synthetic
line observations of molecular clouds \citep{Padoan2016_SN-III} and
for studies of the star-formation rate \citep{Padoan2017_SN-IV}. These
simulations of supernova-driven turbulence were run with the Ramses
code \citep{Teyssier2002}, using a 250\,pc box with periodic boundary
conditions. The runs started with zero velocity, a uniform density
$n{\rm (H)}$=5\,cm$^{-3}$, and a uniform magnetic field of
4.6\,$\mu$G. The self-gravity was turned on after 45\,Myr and the
simulations were then run for another 11\,Myr. In the hierarchical
grid, the largest cell size is 0.25\,pc but in high-density regions
the grid is refined down to $7.6\times 10^{-3}$\,pc. 
In this paper, we use a (10\,pc)$^3$ sub-volume selected from the full
(250\,pc)$^3$ model cloud. Table~\ref{table:levels} lists the number
of cells on each level of this (10\,pc)$^3$ model with maximum
refinement \nL=7. The largest number of cells is found on the level
\nL=4.
Figure~\ref{fig:maps} shows examples of surface brightness maps
computed for models with \nL=1-4.

The radiative transfer problem is solved with SOC, assuming an
external radiation field according to \citet{Mathis1983} (solar
neighbourhood values) and dust properties given by
\citet{Compiegne2011}. We use a fixed frequency grid that has 52
frequencies that are placed logarithmically between 10$^{11}$\,Hz and
$3 \times 10^{15}$\,Hz (between 0.1\,$\mu$m and 3000\,$\mu$m). Tests
are are made assuming that the grains remain in temperature
equilibrium with the radiation field.

Even with external illumination only, the radiation field intensity
does not have significant large-scale gradients in the MHD cloud
models. This is caused by the inhomogeneity of the models which
leads to a relatively uniform intensity in the low-density medium.
Strong temperature variations are seen but mainly at smaller scales,
in connection with individual high-column-density structures. Apart
from the background radiation, the other potential radiation sources
are internal sources (modelled as a blackbody point sources) and the
re-emission from the heated dust itself.

\begin{table}
\centering
\caption[]{Number of cells on each hierarchy level of the
(10\,pc)$^3$ model clouds with \nL=7.}
\label{table:levels}
\begin{tabular}{cr}
\hline \hline
Hierarchy level $L$  &   Number of cells \\
\hline
1  &      8000  \\
2  &     49176  \\
3  &    185232  \\
4  &    365184  \\
5  &    295824  \\
6  &    178280  \\
7  &     83632  \\
\hline
\end{tabular}
\end{table}

\section{Results} \label{sect:results}

To examine the performance of SOC in simulations of external and
internal radiation sources and of dust emission, we start with tests
with simple spherically symmetric model clouds (Sect.~\ref{sect:1D}).
In Sect.~\ref{sect:hires}, we continue with more realistic models 
based on a MHD simulation. Finally, the more demanding iterative
computations with internally heated and optically thick clouds is
discussed in Sect.~\ref{sect:iterations}.

To make the results more concrete, we quote timings for a laptop that
has a six-core CPU and a dedicated GPU\footnote{The CPU is a six-core
Intel Core i7-8700K CPU running at 3.70\,GHz and the GPU an NVidia GTX
1080 with 2560 CUDA cores}. The performance is measured in terms
of the wall-clock time, the actual time that has elapsed, for example,
between the start and the finish of a run.

\subsection{Tests with spherical model clouds} \label{sect:1D}

First tests were conducted with small, spherically symmetric models 
resampled onto a hierarchical grid. The root grid is 17$^3$ cells and
is refined to \nL=2-5 levels with some 5000 cells per level. Thus
about one eight of the cells is refined and the total number of cells
is only $\sim$10000 - 20000, depending on \nL.  A 10000\,K blackbody
point source with a luminosity of $L=1\,L_{\sun}$ is located at the
centre, in a root grid cell with zero density. Otherwise the density
profile is gaussian with a density contrast $\sim$200 between the
centre and the edges. The maximum column density is $N({\rm H_2})\sim
1.3 \times 10^{23}$\,cm$^{-2}$, slightly depending on the
discretisation used. We characterise the noise of the calculations by
using the random mean squared (rms) noise of the resulting 100\,$\mu$m
maps.

\begin{figure*}
%% \sidecaption
\includegraphics[width=17cm]{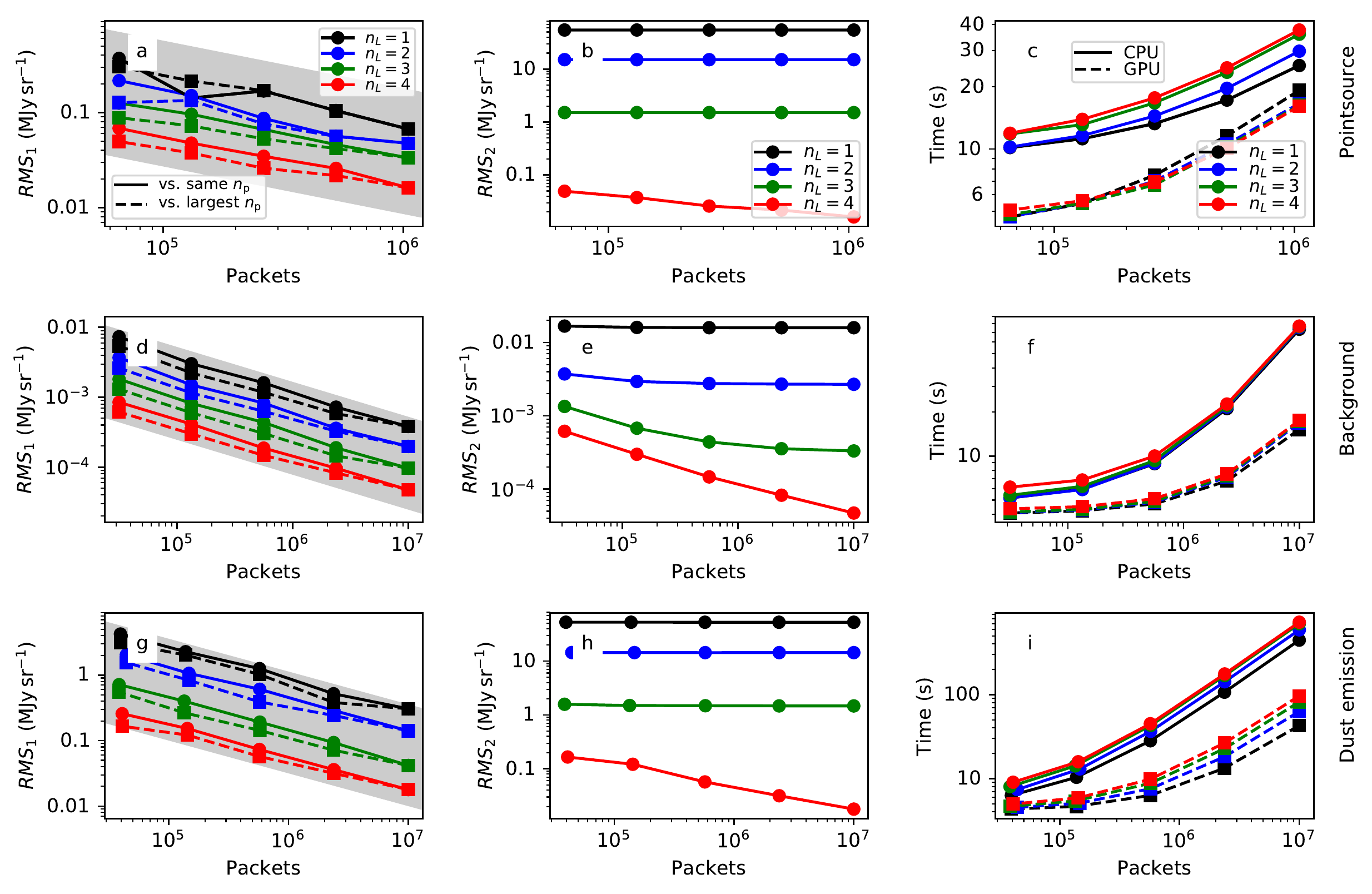}
\caption{
Noise and run times as a function of the number of photon packages for
a small spherically symmetric test model. 
Results are shown for simulations of a point source (frames a-c), a
diffuse background (frames d-f), and the dust emission from the model
volume (frames g-i). 
The first column shows the rms noise of the 100\,$\mu$m map as the
difference of two identical runs (solid lines) and from the comparison
to a run with the same volume discretisation (same \nL) but the
highest number of photon packages (dashed lines). Shading is used to
indicate the expected \nP$^{-1/2}$ convergence as a function of the
number of photon packages.
The second column shows the rms difference to a corresponding run with
the highest number of photon packages $and$ the finest spatial
discretisation (\nL=5).
The wall-clock run times for CPU (solid lines) and GPU (dashed lines)
computations are shown in the last column. In each frame, the colours
correspond to different values of \nL, as indicated in frame c.
}
\label{fig:PS_stat}
\end{figure*}

Figure~\ref{fig:PS_stat} shows how the results change as a function of
the number of simulated photons packages and the model refinement. 
The results show the expected $N^{-0.5}$ dependence of the noise on
the number of photon packages, irrespective of the depth or the grid
hierarchy. This applies to the simulations of the point source, the
diffuse background, and the emission from the medium itself. Each map
has a pixel size that corresponds to the smallest cell size of that
particular model. With larger $L$, the maps become more over-sampled
(especially towards the map edges), which has some effect on the
computed rms values. The actual noise per 3D cell measured by the dust
temperature is in refined regions proportional to $2^L$ because the
number of photon packages hitting a cell decreases by a factor of 4
for each additional level of the grid hierarchy. 

Figure~\ref{fig:PS_stat}b compares results to the map obtained with
the highest refinement and the highest number of photon packages. The
differences are completely dominated by the difference in
discretisation. The dependence on the number of photon packages is
visible only when comparing runs with the same gridding. 

The last column in Fig.~\ref{fig:PS_stat} shows the wall-clock run
times, including initialisations and the writing of the surface
brightness maps for all frequencies. The map size in pixels depends on
the discretisation but the effects for the overall run times are not
significant. On the test computer, the speed-up provided by the GPU
varies from $\sim 50$\% (for small number of packages, when the
initialisation overheads are significant) close to a factor of ten. 
The behaviour is qualitatively similar for lower column density models
where there are fewer multiple scatterings and the cost associated to
the creation of photon package is larger relative to the tracking of
the photon paths. The speed-up of GPU relative to CPU increases
with the number of photon packages, except for the point source
simulations. Possible reasons for this are discussed in
Sect.~\ref{CPUvsGPU}.

%%# OT_ReadHierarchy: `, 20, 20, 20, 7, 1165328)
%%# OT_ReadHierarchy  --- LCELLS[00] =   8000
%%# OT_ReadHierarchy  --- LCELLS[01] =  49176
%%# OT_ReadHierarchy  --- LCELLS[02] = 185232
%%# OT_ReadHierarchy  --- LCELLS[03] = 365184
%%# OT_ReadHierarchy  --- LCELLS[04] = 295824
%%# OT_ReadHierarchy  --- LCELLS[05] = 178280
%%# OT_ReadHierarchy  --- LCELLS[06] =  83632
%% lcells    = [ 8000,49176,185232,365184,295824,178280,83632]
%% tux_cpu   = [ 70.0, 165, 321, 522,  761, 1667 ]
%% gamma_gpu = [   25, 106, 332, 765, 1247, 0    ]

\subsection{Tests with 3D model clouds} \label{sect:hires}

The next tests involve a 3D model based on MHD simulations (see
Sect.~\ref{methods:model}).
%
%It consists of a (10\,pc)${3}$ sub-volume
%that in its densest parts is refined from the 0.488\,pc cell size of
%the root grid down to 7.6\,mpc cells. There are altogether 1.2 million
%cells of which some 84\,000 are on the highest level of refinement. 
%
We analyse maps that are computed towards the three cardinal directions.
The map pixel size corresponds to the smallest model cells. The
10\,pc$\times$10\,pc projected area is thus covered by maps with
1310$\times$1310 pixels. The average column density is 
$N({\rm H}_2)\sim 2.4 \times 10^{21}$\,cm$^{-2}$  ($A_{\rm
V}=2.3$\,mag) and the maximum column density ranges from 
$1.78\times 10^{23}$\,cm$^{-2}$ ($A_{\rm V}=172$\,mag) to
$2.46\times 10^{23}$\,cm$^{-2}$ ($A_{\rm V}=238$\,mag),
depending on the view direction. The values of visual extinction 
$A_{\rm V}$ given in parentheses correspond to the dust model used in
the simulations.

The model volume is centred on a site that will give birth to a
high-mass star. However, in these tests we run the models either
without internal heating or using a radiation source that is located
in the star-forming core but has an ad hoc luminosity that is made so
high that the dust temperatures converge only after several
iterations.

\begin{figure}
%% \sidecaption
\includegraphics[width=8.8cm]{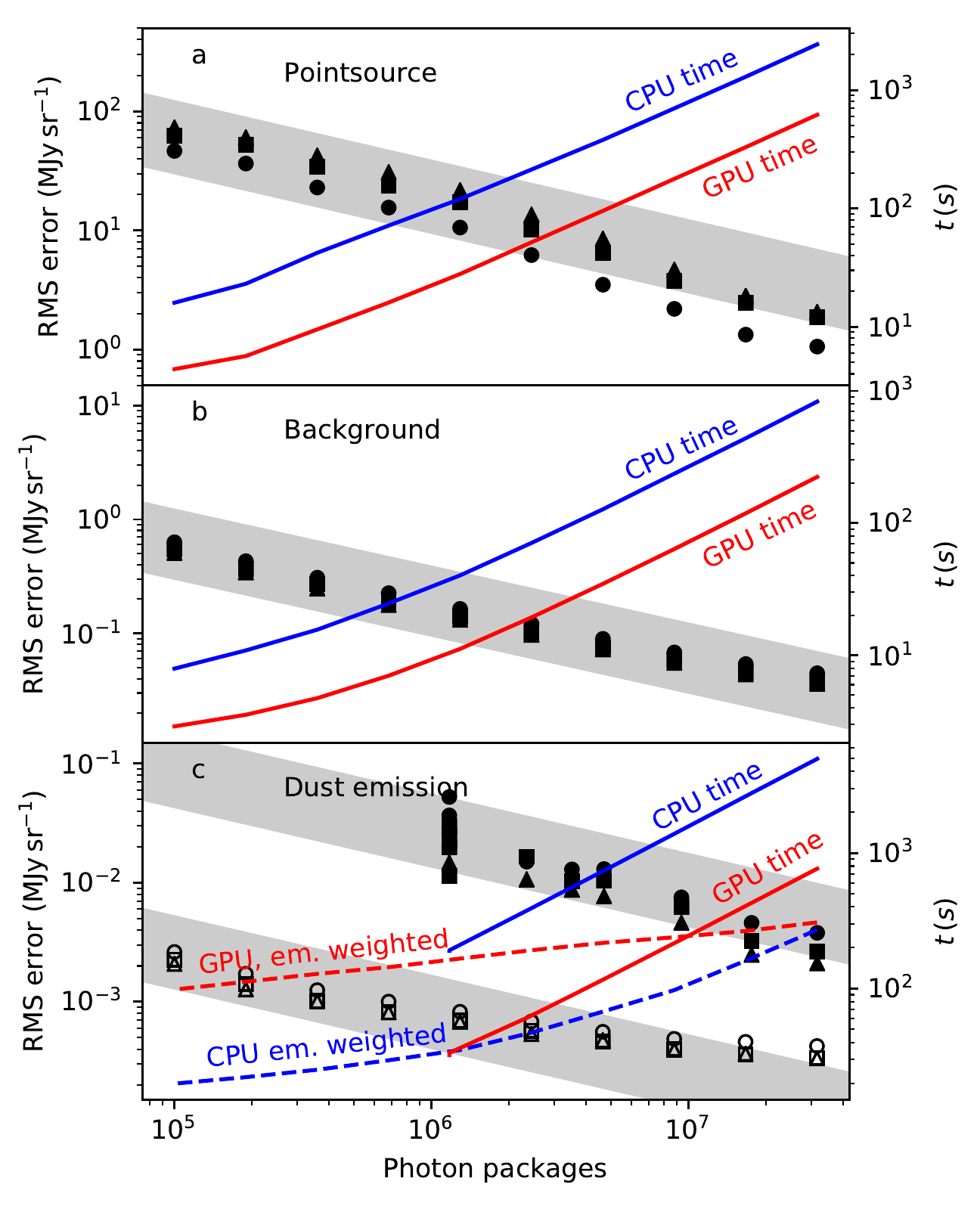}
\caption{
Test with high-resolution model in the case of only a point source
(frame a) or only background radiation (frame b), or including both as
constant radiation field components to examine the noise in the
simulation of dust re-emission. The rms errors as a function of the
number of photon packages are shown with black symbols. The circles,
squares, and triangles correspond to maps calculated toward three
orthogonal directions. The shading shows the expected $N^{-1/2}$
convergence. The lines and right hand y-axis indicate the run times
for CPU (blue) and GPU (red).  In frame c, open symbols and dashed
lines correspond to an alternative run with emission weighting.
}
\label{fig:3Dstat}
\end{figure}

We start by checking how the noise behaves as a function of the number
of photon packages or the run time. Compared to Sect.~\ref{sect:1D},
the grid is now fixed but the model size is closer to that of
potential real applications. Figure~\ref{fig:3Dstat} shows the results
for a single iteration. The noise decreases mostly approximately
according to the $n_{\rm P}^{-1/2}$ relation. In the case of a point
source, the convergence is slightly faster because with lower package
numbers some cells at the model boundaries may not be hit at all.
These cells get assigned an ad hoc constant temperature and, in spite
of their low number, have an impact on overall noise values. 

Plots contain two relations for dust re-emission. In the default
method, the same number of photon packages is sent per each cell (with
random locations and directions within each cell). For uniform
sampling, SOC also requires the number of photon packages (per
frequency) to be a multiple of the number of cells. Therefore, results
for runs with smaller number of {\it requested} photon packages end up
at the same location in the plot with {\it actual} number of photons
packages just above $n_{\rm P}\sim 10^6$. The other relations (open
symbols) correspond to simulations where the number of emitted photon
packages is weighted according to the emission. These runs include, as
constant contributions, the heating from previous simulations of the
external radiation field and the point source. Hot dust (temperature
above 100\,K) is found only near the central source. 
The convergence is slower than $n_{\rm P}^{-1/2}$ because, in the case
of emission weighting, the maximum number of photon packages sent from
any single cell was limited to 10000 with a user-defined parameter.
Such a cap may be sometimes necessary to ensure proper sampling also
for the emission from cooler dust, which could be important locally 
in regions far from the hottest dust. On the other hand, an increase
in $n_{\rm P}$ will not improve the accuracy with which one simulates
the emission from cells that have already reached the cap value.

Apart from constant overheads at small photon numbers, the run times
are generally directly proportional to \nP. The speed-up provided by
the GPU is $\sim$4 in the pointsource and background radiation
simulations and slightly higher for the standard dust re-emission
calculation. The situation is very different when emission
weighting is used. Compared to the unweighted case, the run times are
shorter on CPU while on GPU they are longer. This is discussed further
in Sect.~\ref{CPUvsGPU}.

\begin{figure}
%% \sidecaption
\includegraphics[width=8.8cm]{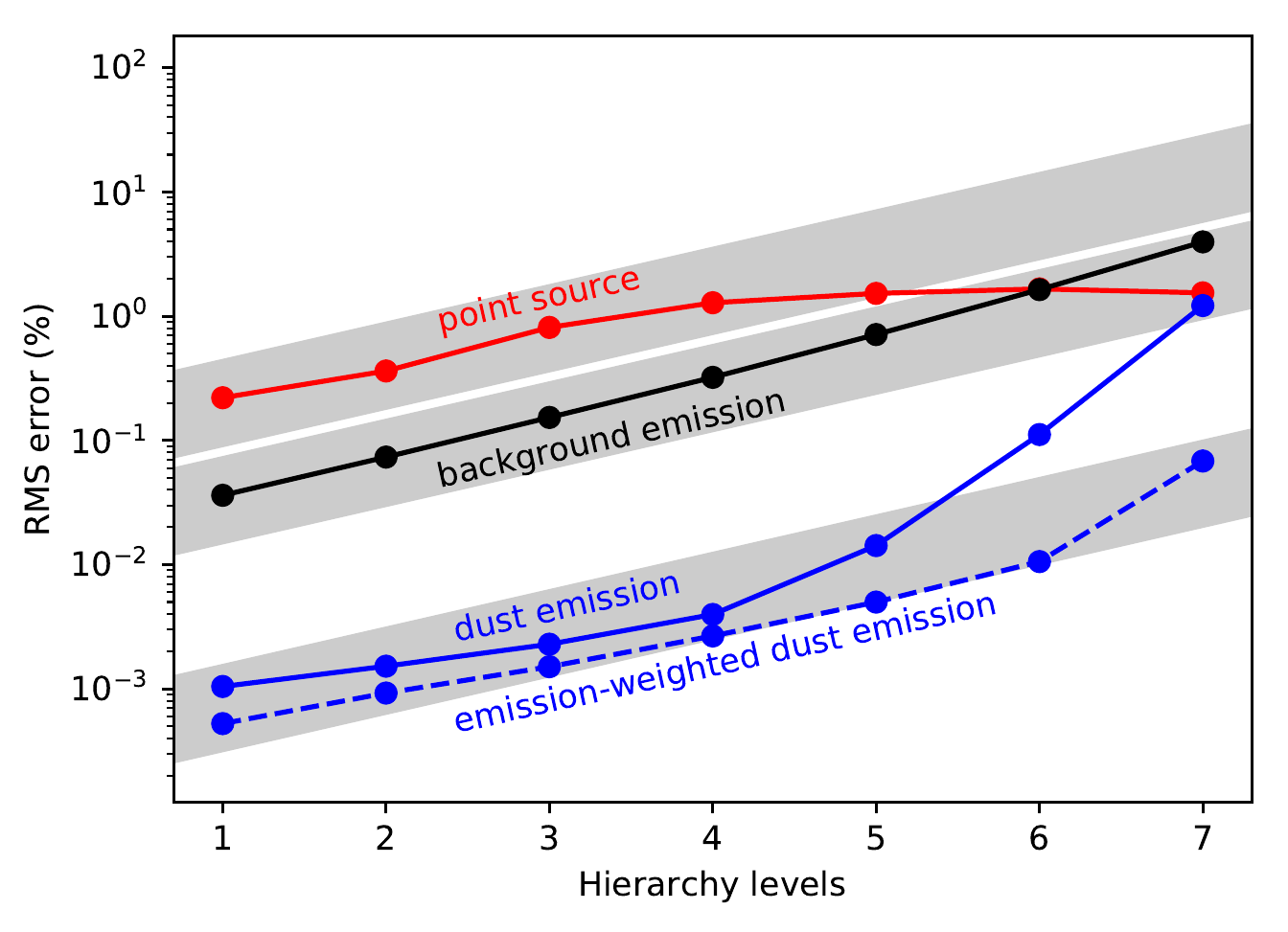}
\caption{
Error of the computed dust temperatures as a function of the grid
hierarchy in a model with \nL=7. From top to bottom, the curves
correspond to simulations of a point source (red curve), heating by
background radiation (black curve), and simulations of dust emission
when the contribution of the previous radiation sources is kept
constant (blue curves). The solid and dashed lines correspond to the
default and the emission-weighted simulations, respectively. The
shaded regions are used to illustrate the expected $2^{\rm level}$
dependence of the noise. All calculations were done with $1.8 \times
10^7$ photon packages per frequency.
}
\label{fig:dT_vs_level}
\end{figure}

Without photon splitting or similar techniques, each additional level
of the hierarchy should decrease the probability of photon hits by a
factor of four. Thus, the noise should increase proportionally to
$2^L$, where $L$ is the grid level (larger $L$ standing for smaller
cells). Figure~\ref{fig:dT_vs_level} shows the actual rms error of 
dust temperatures as a function of the hierarchy level. The
$2^L$-dependence holds for the background radiation. In the
pointsources simulation, the noise is constant or even decreases at
the highest levels because those cells are on average closer to the
source. For the simulated dust re-emission, the noise values are lower
but increase rapidly on the highest refinement levels. This is partly
an artefact of the model setup where the cells on low grid levels are
heated mainly by sources other than the dust re-emission and therefore
in this test have a low noise. The emission-weighting leads to lower
noise values that also are more uniform between cells of different
size.

\begin{figure}
%% \sidecaption
\includegraphics[width=8.8cm]{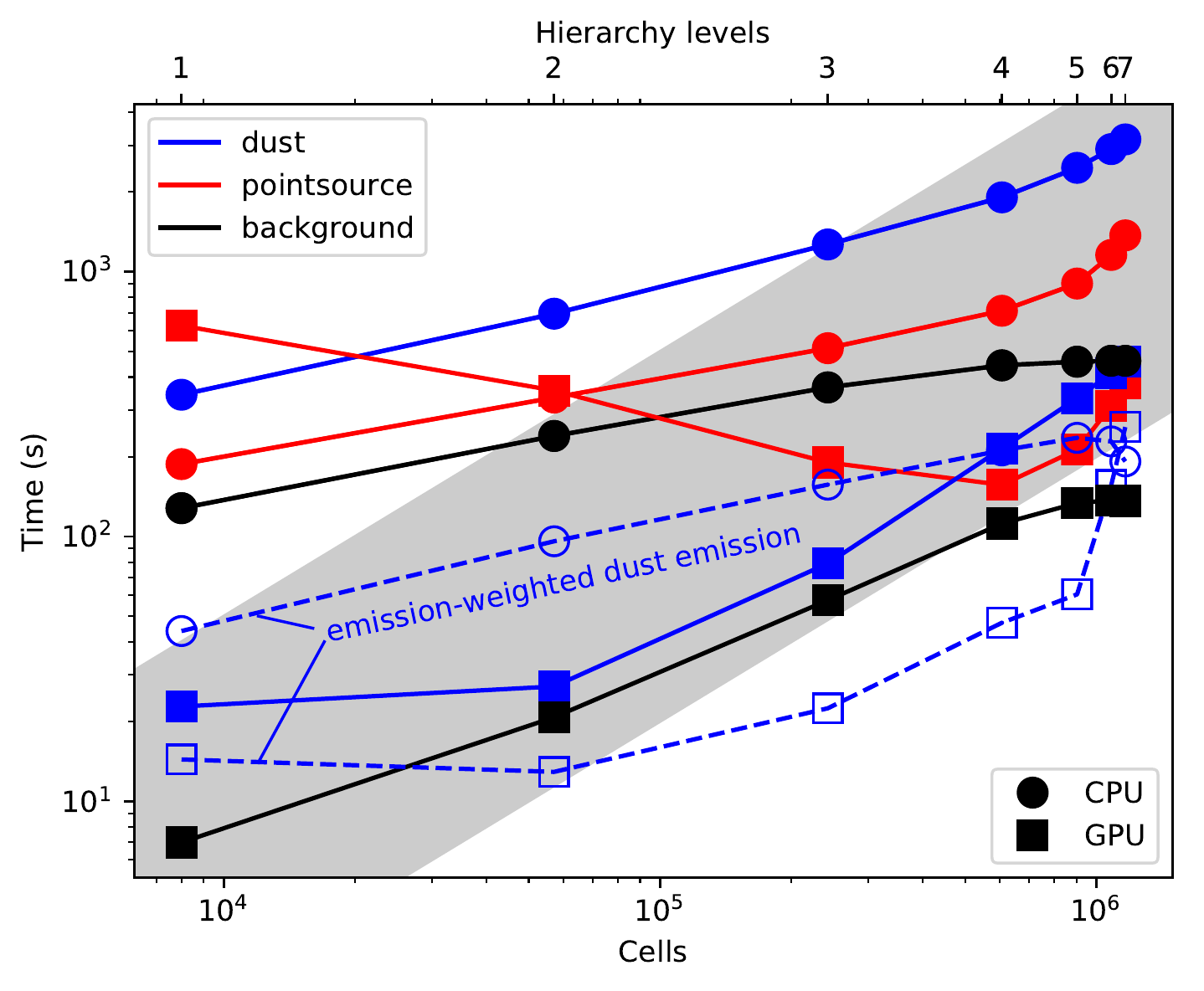}
\caption{
Dependence of run times on the maximum refinement. The colours
correspond to simulations of the radiation from a point source (red),
external background (black), and dust re-emission (blue). For dust
emission, the dashed lines and open symbols correspond to a run using
emission weighting. The CPU run times are plotted with circles and the
GPU run times with squares. The shading indicates the trend for
run times proportional to the number of cells.
}
\label{fig:time_vs_cells}
\end{figure}

Run times should be proportional to the number of cells (assuming
that the sampling of the radiation field is kept uniform) but, in our
case, deeper grid hierarchies are associated with some overhead in the
tracking of the photon packages. Figure~\ref{fig:time_vs_cells} shows
the run times as a function of the number of cells in models that are
refined down to \nL=1-7. The number of photon packages is kept at $1.8
\times 10^7$ so that plot shows only the effect of discretisation. 
In the plot, the run times increase slower than the number of cells
because cells at higher hierarchy levels are hit by progressively
fewer photon packages. To ensure a certain SNR level even for the
refined regions, the number of photon packages should be proportional
to $2^{n_{\rm L}}$ (Fig.~\ref{fig:dT_vs_level}). An increase in the
overhead is visible for deeper hierarchies with \nL=5-7 although the
run times are still almost proportional to the number of cells.  As
indicated by Table~\ref{table:levels}, even though the tracing of the
photon paths through cells at high refinement levels were
significantly slower, the effect on the total run times is limited
because of the small fractional number of those cells. Some GPU runs
appear to deviate from the general trends. Emission-weighted
simulations of dust re-emission slow down for deeper hierarchies (see
Fig.~\ref{fig:3Dstat} and discussion above) and for \nL=5-7 are
similar to the CPU run times. Furthermore, point source simulations
suffer from coarse discretisation (\nL$<$=4), possibly because of the 
locking overheads for global (multiple threads doing frequent
updates to the same cells close to the point source).

\subsection{Iterative solution of dust temperature} \label{sect:iterations}

Above we were only concerned with the simulations of the radiation
field. For optically thick models and especially in the case of
internal radiation sources, the question of the convergence of the
temperature values becomes equally important. Calculations could be
sped up by reducing the number of iterations required for a converged
solution or by reducing the time per iteration. 

We first examine the performance of the ALI method as a function of
the model discretisation. We are using ALI with a diagonal operator
that only separates the absorption-emission cycles within individual
cells. The effects should depend on the discretisation. Higher spatial
resolution means that the optical depth of individual cells is smaller
and thus we can expect less benefit from the use of ALI. 

We use the same density field as in Sect.~\ref{sect:hires} and a
single internal radiation source. The original model described a
(10\,pc)$^{3}$ volume where the effects of an even very luminous
source would remain local. Therefore, in this test the linear size of
the cloud was decreased by a factor of 20, the densities were
increased by a factor of 100, and the luminosity of the radiation
source set so that the temperature of the closest cells is hundreds of
degrees. These ensure that dust re-emission is important over a larger
volume and that the dust temperatures converge only after many
iterations. The setup is only used to test the radiative transfer
methods and is not supposed to be a physically accurate description of
a star-forming core.

\begin{figure}
%% \sidecaption
\includegraphics[width=8.8cm]{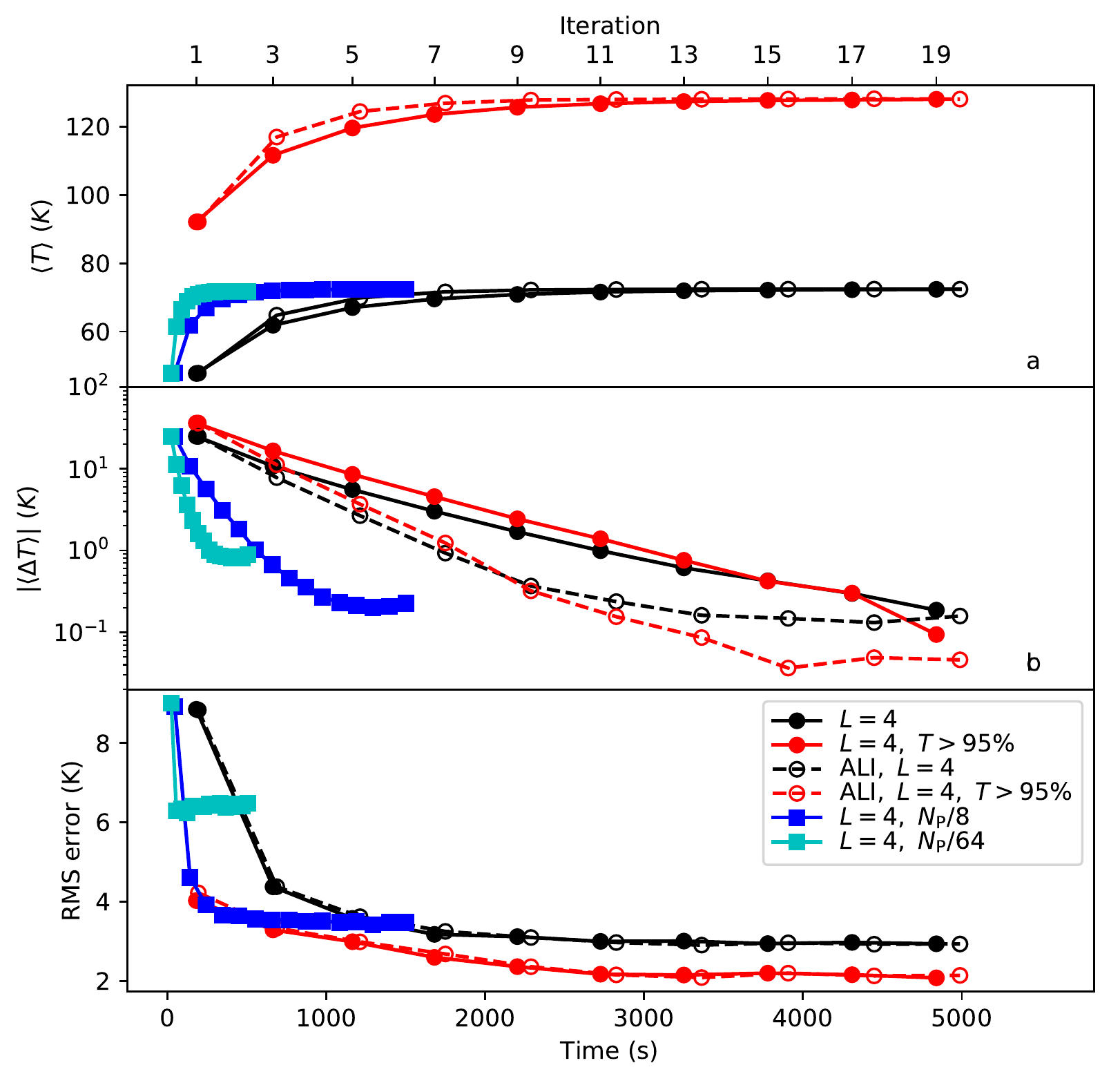}
\caption{
Convergence of dust temperature values as a function of iterations and
the run time on CPU. Temperatures are from the cells on the finest
level of refinement of a model with \nL=4. Frame a shows the average
temperature for basic runs (solid lines) and for ALI runs (dashed
lines and open symbols). Black line is the average for all cells on
the level $L$=4 and the red curve is for 5\% of the cells with the
highest temperatures. The blue and cyan curves correspond to the
average of all $L=$4 cells in runs with reduced number of photon
packages \nP.
%%%
Frame b shows the errors relative to a run with 40 iterations and
without ALI. Frame c shows the rms errors for $L=4$ cells calculated
as the standard deviation between two independent runs.
}
\label{fig:T_vs_iter}
\end{figure}

Figure~\ref{fig:T_vs_iter} shows the convergence of temperature values
for the \nL=4 model for runs with \nP$\sim 18\times 10^6$.  The
convergence is measured based on the average dust temperatures on only
the highest level of refinement (thus cells close to the point
source), comparing these to a non-ALI run with the same \nP\, and 40
iterations. Based on the rate of convergence in
Fig.~\ref{fig:T_vs_iter}, the error of that reference should be two
orders of magnitude smaller than on the final iterations shown in the
figure.

The average temperature is initially increasing by $\sim$10\,K per
iteration. After 20 iterations, this rate has decreased by a factor of
one hundred.  ALI leads to a faster convergence and the rate of
convergence is about the same for all cells, irrespective of their
location with respect to the point source.
In run times, the overhead of ALI is some 5\%, which is more than
compensated by the faster convergence. The rms errors are similar with
and without ALI.  

Figure~\ref{fig:T_vs_iter} shows further how, depending on the
requirements on accuracy (bias and random errors), the run times could
be significantly shortened simply by reducing the number of photon
packages. The rms errors are naturally higher but this does not affect
the initial rate of convergence measured using $|\langle \Delta T
\rangle|$. However, with low photon numbers, the convergence stops 
earlier and also the bias of the final temperature estimates is
larger. When \nP\, is reduced by a factor of 64, the final bias is
$\Delta \langle T \rangle \sim$1\,K. The bias is even more significant
in the resulting surface brightness maps, analogous to the way LOS
temperature variation bias observational dust temperature estimates
\citep{Shetty2009a, Juvela2012_Tmix}. Figure~\ref{fig:T_vs_iter}. 

In Fig.~\ref{fig:T_vs_iter}b the convergence of the ALI runs saturates
at a level $|\langle \Delta T \rangle|\sim0.1$\,K. Qualitatively this
could be expected based on the above results with lower photon number.
However, we measure convergence with respect to a reference solution
that is calculated with 40 iterations but without ALI. Like the
low-photon-number runs, the reference solution will have a systematic
error that is larger than suggested by the extrapolation of the
initial linear trend in Fig.~\ref{fig:T_vs_iter}b. The fact that the 
$|\langle \Delta T \rangle|$ curve of the ALI run flattens relative to
that reference solution suggests that ALI may be more sensitive to
Monte Carlo noise.

The emission maps and error maps for the final iteration of
Fig.~\ref{fig:T_vs_iter} are shown in Appendix~\ref{sect:maps}.

\subsection{Faster iterative solutions} \label{sect:accel}

When the convergence of temperature values requires many iterations,
the calculations can be sped up in at least three ways. These include
the use of sub-iterations, the use of a reference field, and run-time
model refinement.

The idea behind sub-iterations is that temperature updates are
sometimes restricted to an ``active'' subset of cells whose
temperatures and temperature changes are most relevant for the final
solution. This results in savings in the simulation step, because
emission is re-simulated only from a fraction of all cells, and in the
temperature updates, which are similarly restricted to a subset of
cells. Some overhead is caused by the necessity to use a reference
field to store the contribution of other cells to the total radiation
field. We do not test this method here because emission weighting
provides similar savings in the simulation step. Because our tests do
not include stochastically heated grains, the contribution of
temperature updates to the total run times is at the level of only one
per cent. For stochastically heated dust, the run times could be
dominated by the calculation of the temperature distributions 
(see Appendix~\ref{sect:SHG}) and a simpler version of sub-iterations
can be implemented simply by skipping the temperature updates for
weakly-emitting cells.
 
The use of a reference field enables speed-up because, for a given
final noise level of the solution, the number of photon packages per
iteration can be lower. In the case of a hierarchical grid, the most
refined regions tend to have both the highest noise and the slowest
convergence. Therefore, some care must be taken to ensure that the
solution has truly converged in those regions.

\begin{figure}
%% \sidecaption
\includegraphics[width=8.8cm]{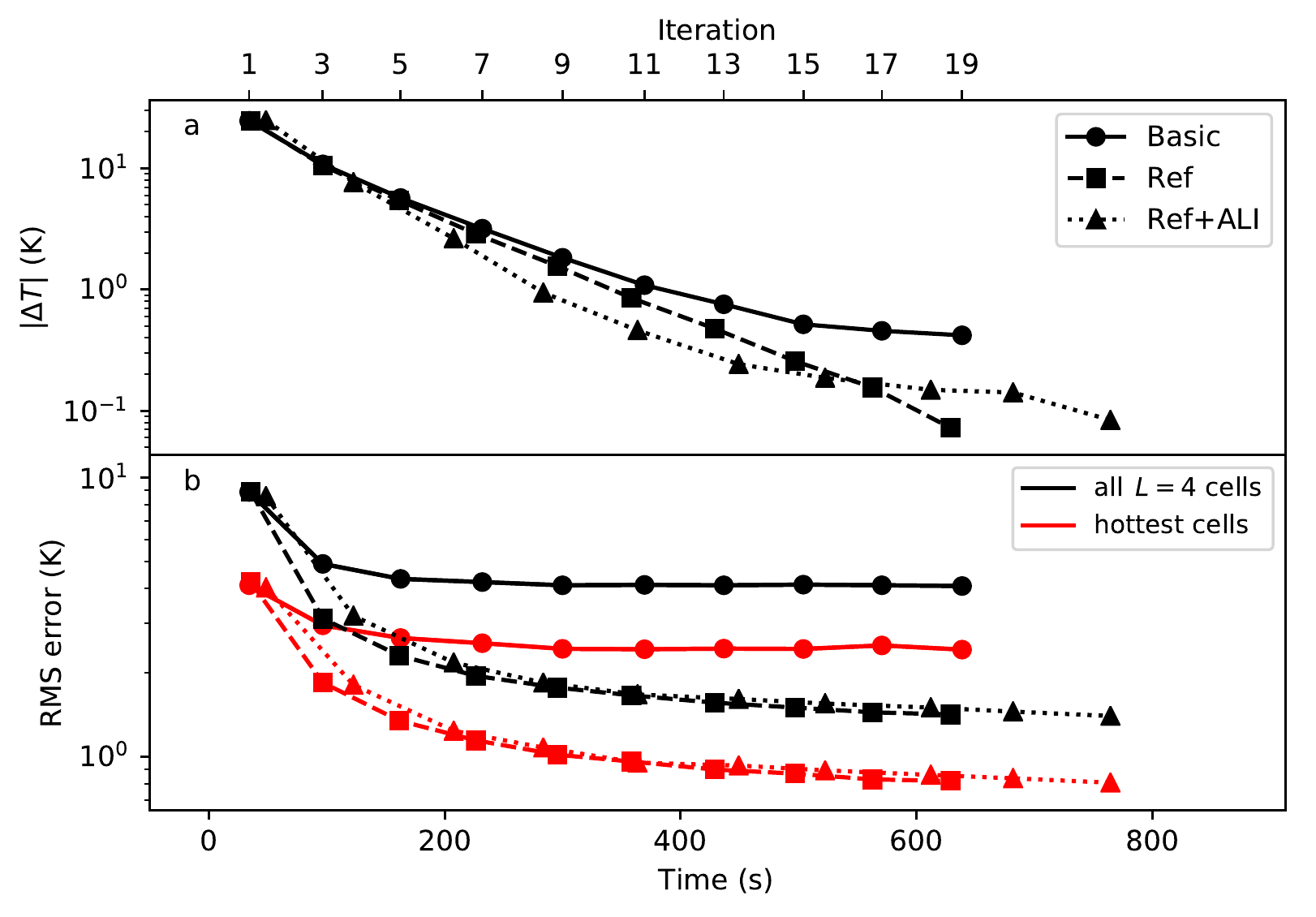}
\caption{
Results for \nL=4 model when using a reference field.  All statistics
refer to the average dust temperature of the cells on the highest
$L$=4 level of the grid hierarchy. 
%%
%% Frame a shows the convergence of the average temperature and 
Frame a shows the temperature error relative to the solution obtained
with the same setup after 20 iterations.
Frame b shows the rms temperature difference between two independent
runs, the red curves showing these separately for 5\% of the cells
with the highest temperatures.
Circles stand for the basic calculations (no ALI and no reference
field), squares to a reference field runs, and triangles to the
combination of the reference field and ALI techniques.
Values are plotted against the run time and the upper x-axis shows the
corresponding number of iterations in the basic run. 
}
\label{fig:plot_RF}
\end{figure}

Figure~\ref{fig:plot_RF} shows results for the \nL=4 model when using
a reference field. This can be compared to the previous plots of the
rms noise in Fig.~\ref{fig:dT_vs_level} and the convergence in
Fig.~\ref{fig:T_vs_iter}. The setup is identical (including background
and a point source and constant radiation sources) but the number of
photon packages per iteration has been decreased by a factor of 16.
Examples of the corresponding surface brightness maps can be
found in Appendix~\ref{sect:maps}.

The run times are significantly shorter than in
Fig.~\ref{fig:T_vs_iter} but only by a factor of $\sim 6$ rather than
the factor of 16 between the number of photon packages. This is caused
by the non-linearity seen in Fig.~\ref{fig:PS_stat}i, the lower
efficiency of simulations with low photon numbers that should
disappear for sufficiently large photon numbers. The convergence rate
in Fig.~\ref{fig:plot_RF}b should depend only on the use of ALI. ALI
results in smaller $\Delta T$ errors but the difference is smaller
than in Fig.~\ref{fig:T_vs_iter}b, probably because the larger noise
on the first iteration decreases the efficiency of ALI corrections.
The effect of a reference field is seen in Fig.~\ref{fig:plot_RF}b
where it decreases that final rms error by a factor of $\sim 3$. The
combination of ALI and a reference field increases the run times per
iteration by close to 30\% while the reference field alone is not
associated with any overheads.

%%%%%%%%%%%%%%%%%%%%%%%%%%%%%%%%%%%%%%%%%%%%%%%%%%%%%%%%%%%%%%%%%%%

\subsection{Run-time grid refinement}

\begin{figure}
%% \sidecaption
\includegraphics[width=8.8cm]{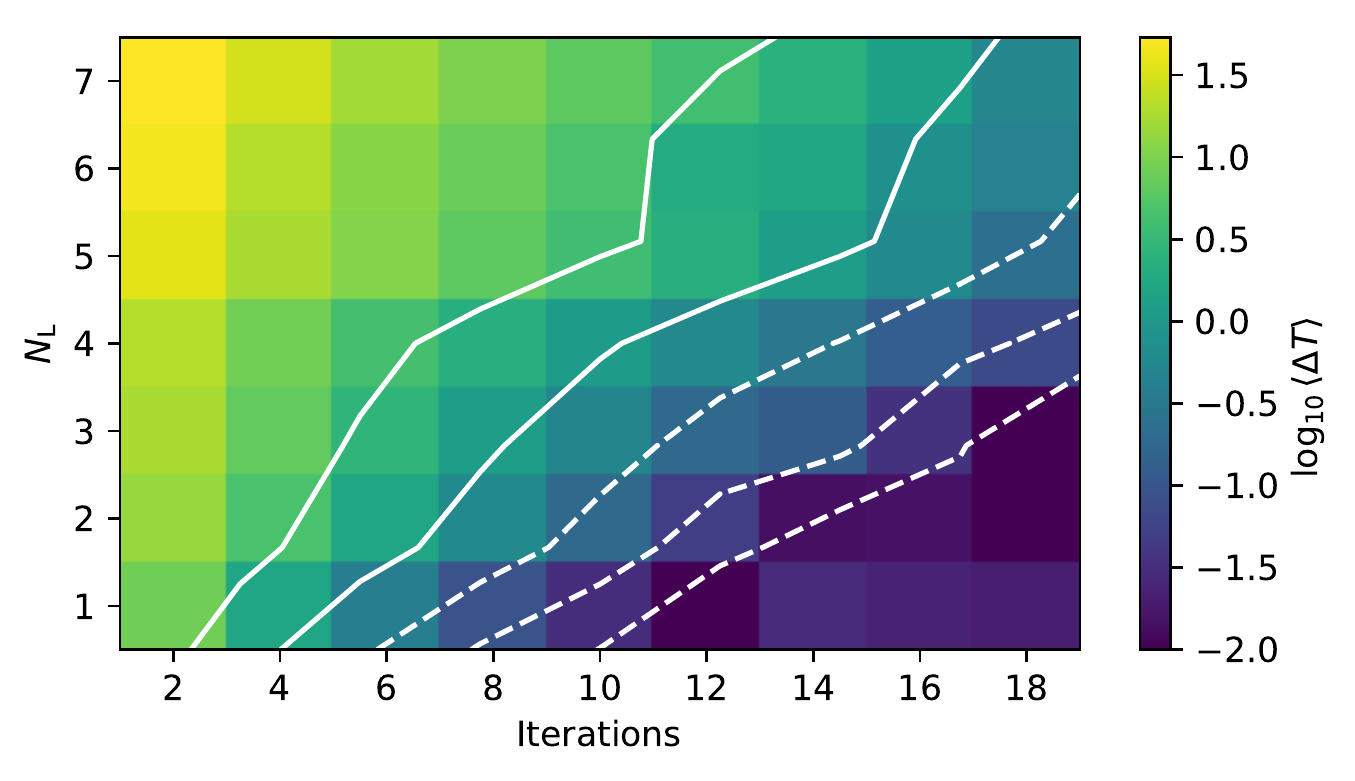}
\caption{
Convergence of dust temperature values for models of different
refinement. The colour scale shows the difference of the average dust
temperature at the highest level of refinement ($L$=\nL) compared to
the solution after 20 iterations. The number of hierarchy levels \nL
is on the y-axis. Calculations are done without ALI and without a
reference field. The contours are drawn at ${\rm log}_{10} \,a$ for
$a$=-1.5, -1.0, -0.5, 0.0, and 0.5.
}
\label{fig:dT_map}
\end{figure}

Before discussing tests with run-time model refinement,
Fig.~\ref{fig:dT_map} shows for how dust temperature errors depend on 
the number iterations and the number of hierarchy levels (\nL). 
Neither ALI not a reference field is being used and the discretisation
is constant throughout the calculations that each correspond to one
row in the plot. The plot only measures the errors related to the
convergence of temperature values. The errors are calculated for cells
at the highest level ($L=$\nL), in relation to the result obtained
with the same \nL after 20 iterations. Discretisation errors are
therefore excluded from the plot. The figure quantifies how an
approximate solution of a lower-resolution model can be found with
fewer iterations. For example, an error of $\Delta T= 0.1$\,K (${\rm
log}_{10}\,\Delta T=-1$) is for \nL=1 reached in 8 iterations while
$\sim$20 iterations are needed for \nL=4. The full-resolution model
would need at least 30 iterations. For the idea of run-time grid
refinement, this means that the initial iterations with low-resolution
models are not only faster (time per iteration) but their number also
should remain small compared to the total number of iteration.

\begin{figure}
%% \sidecaption
\includegraphics[width=8.8cm]{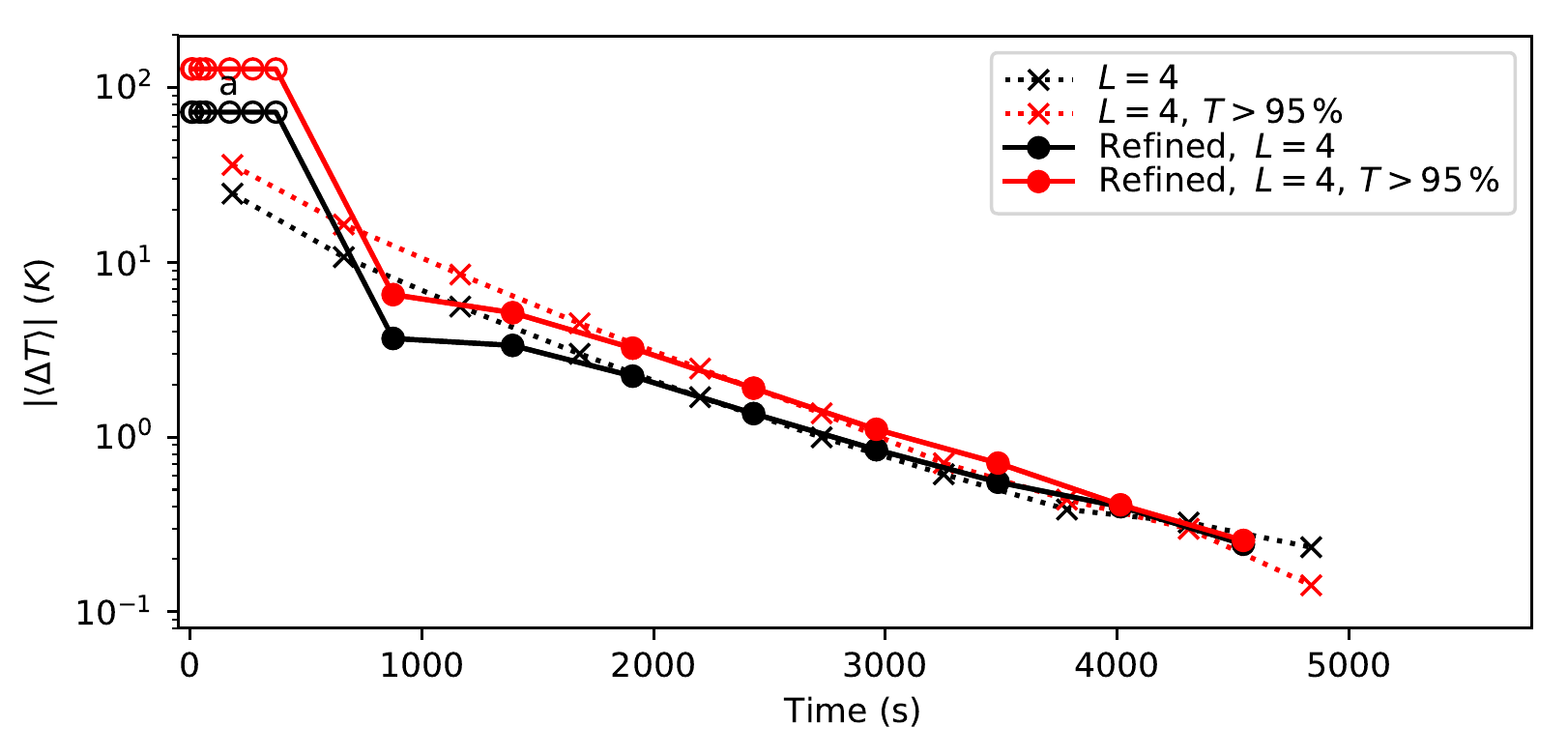}
\caption{
Convergence of temperatures for basic runs with the \nL$=4$
model(dashed lines) and with run-time refinement of the spatial grid
from \nL$=1$ to \nL$=4$ (solid lines).  The results are shown for the
average of all $L=4$ cells (black lines) and for the average of 5\% of
the warmest $L=4$ cells (red lines). 
Convergence $| \Delta T |$ is measured as the difference to a run with
40 iterations and the same \nP. In the case of run-time refinement,
the open symbols corresponds to initial iterations with \nL$<4$, are
drawn at $| \Delta T |= 100$\,K, and only indicate the run times of
those steps.
}
\label{fig:plot_LEV}
\end{figure}

Figure~\ref{fig:plot_LEV} shows the results for a run with run-time
refinement of the model. Calculations start with the root grid only
(\nL=1). One level is added on iterations 5, 9, and 15, thereafter the
remaining run proceeds with the full \nL=4 model. When the grid is
refined, also the number of photon packages is quadrupled so that the
final iterations have the same 18 million photon packages per
frequency as in previous tests (e.g. Fig.~\ref{fig:dT_vs_level}). 
%% ALI and reference field methods were not used. 

The initial iterations are very fast. However, 
Fig.~\ref{fig:plot_LEV} shows that run-time refinement does not have a
significant effect on the final convergence. After the final
refinement, the error $|\Delta T|$ is half of the value of the basic
run that corresponds to the same run time. However, subsequent
convergence remains almost identical to the case where the \nL$=4$
grid was used on all iterations.

\subsection{Variable number of photon packages}

Apart from the number of cells, the number of photon packages is the
main factor determining the run times. Therefore, we investigated if a
good solution can be found faster if the number of packages was
increased during the iterations. An acceptable solution clearly
requires both convergence and low random errors.
Figure~\ref{fig:T_vs_iter} already suggested that these are not
independent and systematic error measured by $\Delta T$ also depend on
the random Monte Carlo noise.

\begin{figure}
%% \sidecaption
\includegraphics[width=8.8cm]{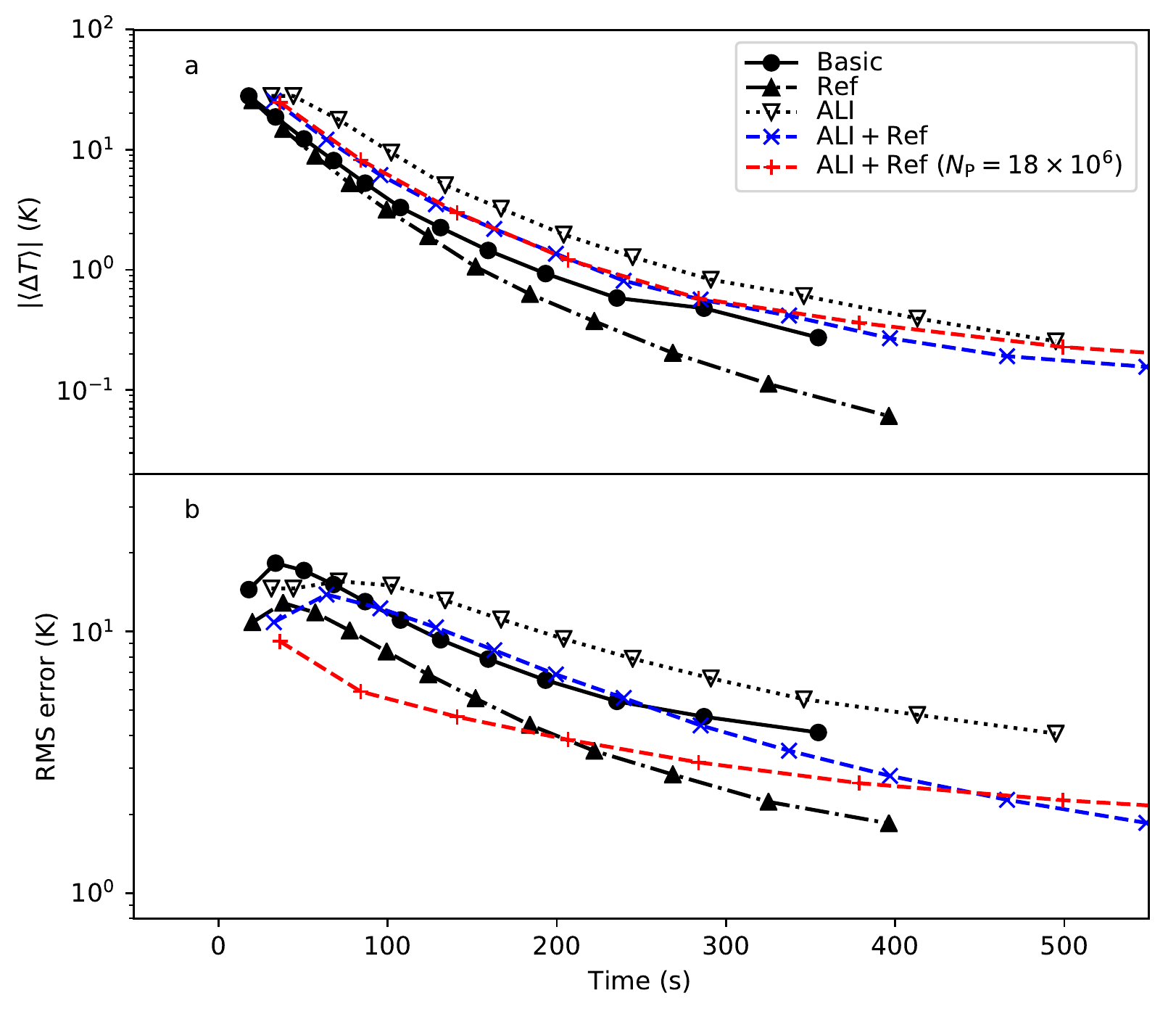}
\caption{
Convergence and rms errors of temperatures in runs where the number of
photon packages \nP\, is increased by a factor of 200 over the
iterations. The final iterations use 18 million photon packages per
frequency. Results are shown for different combinations of ALI and
reference field methods, as indicated in frame a. For reference, the
red curves show the results for the constant \nP\, of 18 million
photon packages per iteration. Markers are drawn for every second
iteration.
}
\label{fig:plot_PN}
\end{figure}

In Fig.~\ref{fig:plot_PN} we examine runs with 22 iterations where the
final number of photon packages \nP\, is again 18 million. However, this
time the initial number of photon packages is smaller by a factor of
200 and photon numbers are increased so that $log \, n_{\rm P}$ grows
linearly over the iterations.

For the basic method and the reference field method the rate of
convergence is similar but the final rms error of the reference field
method is two times lower. In principle, if \nP\, were constant, the
reference field method could result in an rms noise that is smaller by
a factor of $N_{\rm I}^{-1/2}$. The advantage is here smaller because
initial iterations employ a smaller number of photon packages. More
importantly, the reference field is less efficient because the dust
temperatures change significantly over the iterations.

Unlike for example in Fig.~\ref{fig:T_vs_iter}, the use of ALI not
only increases the rms errors but also results in a slower
convergence. This is probably a result of large Monte Carlo noise on
the initial iterations that also renders the ALI updates noisy and
thus ineffective. The combination of ALI and a reference field fairs
somewhat better but the convergence measured against the run time is
still worse than without ALI. For reference, Fig.
Fig.~\ref{fig:T_vs_iter} also shows the result for the combination of
ALI and a reference field when all iterations use 18 million photon
packages. There the convergence per iteration is much faster.  When
measured against the actual run time, the convergence in $|\Delta T|$
is initially similar and later inferior to the variable $|n_{\rm P}|$
runs.

\section{Discussion}  \label{sect:discussion}

In this paper, we have described the implementation of the radiative
transfer programme SOC and quantified its performance in the modelling
of dust emission from interstellar clouds. Unlike most benchmark
papers, we also examined the actual run times and especially the
relative performance of CPUs and GPUs.

\subsection{Radiative transfer methods} \label{methods}

SOC is based on Monte Carlo simulations and the tests showed that
it behaved according to expectations. The Monte Carlo noise is
inversely proportional to the square root of the number of photon
packages. Similarly, the noise is dependent on the level of the grid
refinement, each additional hierarchy level approximately decreasing
the number of photon package hits by a factor of four and increasing
the rms noise by a factor of two. The experiments with photon package
splitting schemes were mentioned in this paper only briefly because
their success was found to be limited. They do decrease somewhat the
noise at higher refinement levels and might be indispensable for
deeper hierarchies. However, in our tests there were no significant
savings in the actual computational cost. These methods need to be
investigated further. The penetration of external radiation into
high-density regions is decreased by backscattering and, therefore,
photon splitting might need to be combined with directional weighting
to increase the probability of photon packages reaching the highest
column density regions. 

In addition to the pure radiation field simulations, we studied the
convergence of the dust temperatures in optically thick models with
internal heating sources. The ALI and reference field methods were
tested, also in relation to the other run parameters like the number
of photon packages per iteration and the discretisation of the model
volume.

ALI improves the temperature convergence but is relevant only
when both optical depths and dust temperatures are high. In the
context of interstellar clouds, benefits are noticeable only near
embedded heating sources, potentially in a very small fraction of the
whole model volume (although in relative terms a larger fraction of
grid cells). To reduce memory requirements (one additional number per
cell) and run times, the use of ALI could be restricted to selected
cells. This would be only a small complication in the RT
implementation. However, in our tests where ALI was included for all
cells, the run time overhead was not very significant, only some tens
of percent.

The most straightforward way to speed calculations is to use fewer
photon packages. This will of course increase the noise of the
solution but, when solution also requires a number of iterations, 
gives room for further optimisations. Significant savings are possible
if initial iterations are done fast, with low number of photon
packages \nP. Parameter \nP\, can be increased with iterations, to
keep random errors comparable to the convergence errors. This turned
out to work less reliably in connection with ALI, which provided clear
acceleration only when the random errors were kept low. The final rms
error of the solution also tended to be larger when ALI was used
(Figs.~\ref{fig:plot_RF}, ~\ref{fig:plot_PN}). 

The most robust way to speed-up computations was to gradually increase
\nP, without ALI but using the reference field method
(Fig.~\ref{fig:plot_PN}).  Without a reference field, the final
accuracy will be limited by the number of photon packages per
iteration (clearly demonstrated by Fig.~\ref{fig:T_vs_iter}). This
determines the rms errors but may also impact the convergence. With a
reference field, the rms errors decrease as iterations progress,
making it also easier to track the convergence. The fastest
convergence in terms of the run time was reached when the number of
photon packages \nP\, is increased with iterations. Changes in \nP\,
must of course be taken into account also in the reference field
updates. Initial iterations are thus given a smaller weight, which is
also useful because the reference field is still far from the final
solution.

ALI methods are very likely still necessary for models more optically
thick than the ones tested in this paper. The critical quantity is the
optical depth of individual cells, especially if the ALI scheme does
not explicitly consider longer-distance interactions. In general,
hierarchical grids enable finer discretisation of dense regions, which
reduces the optical depth of individual cells and thus reduces
benefits of using ALI. In this paper, only the diagonal part of the
$\Lambda$ operator was separated. The next step, explicit treatment of
the radiative couplings between a cell and its six Cartesian
neighbours, may already be excluded by the associated storage
overhead. The implementation in connection with a hierarchical grid
would also be significantly more complicated. 

Finally, we also tested runs where the spatial discretisation was
refined during the iterations. A large cell size used on the first
iterations could increase the speed with which information traverses
an optically thick model, especially if ALI is used to accelerate the
convergence within the individual cells. In practice, the results were
not very encouraging (Fig.~\ref{fig:plot_LEV}). Iterations with low
\nL are indeed very fast but changes in the gridding cause large
temperature jumps and the final convergence is not improved. In the
tested model, the temperature field contains significant structure
even at very small scales. A change in discretisation significantly
changes the radiative transfer problem itself (cf.
Fig.~\ref{fig:PS_stat}), also changing the optical depths between the
radiation sources and the cells. Our implementation of the re-gridding
procedure could be improved. We assigned the temperature of a parent
cell to all its children while 3D interpolated values should work
better. In smooth models with small density and radiation field
gradients, the convergence would naturally be less disturbed by
changes in the grid.

We tested SOC using calculations with a single dust population and
assuming that the grains are at an equilibrium temperature.  SOC will
be developed further to allow the modelling of multiple dust
components. This requires only small kernel modifications but implies
an increase in the memory usage. If the scattering function was
constant, one could store for each cell (instead of one number, the
density) both the absorption and scattering cross sections. However,
to model the scattering accurately, the abundance of each dust
population is needed separately for each cell. This makes it possible
to make run-time Monte Carlo sampling from the ensemble of scattering
functions. However, this will also at least double the memory usage. 

Full radiative transfer modelling of dust emission requires both the
evaluation of the radiation field and the estimation of the resulting
dust emission. When grains are assumed to remain at an equilibrium
temperature, the latter problem is not significant for the overall
computational cost. However, to model mid-infrared emission from
stochastically heated grains, the run times may become dominated by
this task, which includes the computation of temperature probability
density distributions for each cell, grain population, and grain size.
SOC delegates this to an external program. However, in
Appendix~\ref{sect:SHG} we discuss how also these calculations can be
sped up by using GPU computing. The use GPUs in this context was
already discussed, for example, in \citet{Siebenmorgen2012}.

\subsection{Implementation and CPU vs. GPU comparisons} \label{CPUvsGPU}

The current SOC version is implemented in Python that is convenient
for rapid development but, as an interpreted language, is not expected
to be fast. In practice, this is not a problem because most
calculations are performed by the OpenCL kernels that are compiled at
run time. In CPU runs with 3D models (\nL=2), some 93\% of the total
run time was used by the kernel for the radiative transfer simulation,
less than 1\% by the dust temperature calculations, and some 2.5\% by
the kernel for the map calculations. The remaining part includes the
execution of the Python host code, the compilation of the OpenCL
kernels, the data transfer between the host and the device, and the
reading and writing of data files. In CPU runs these thus amount to
little more than 3\% of the total run time but, because of the much
faster kernel execution, can in short GPU runs reach 30\%. However,
this would not disappear entirely even if the host code were compiled.
Indeed, comparisons with previous SOC versions, where the main program
was written in C++, suggested at most $\sim$10\% overhead due to the
use of Python.

Experience has shown that SOC is usually (but not always) slightly
faster than CRT \citep{Juvela2005}, when both are run using CPUs and 
the same non-hierarchical grids. In CRT the critical parts have been
parallelised using OpenMP. However, comparisons are not trivial
because of slight differences in the implementations (e.g., how
randomness is reduced in case of different radiation sources or how
the run times scale with the model size). Such tests would be useful,
but only when conducted systematically over a large number of test
cases and comparing the actual noise properties of the results.

Theoretical peak performance of modern GPUs is very high, which could
translate to much shorter run times and smaller energy consumption
\citep{Huang2009, Said2015}. In the test system the theoretical ratio
was over a factor of 20 in favour of the GPU. In practice the
theoretical speed-up is never reached, mainly because of data transfer
overheads. This was true also in the SOC tests where the GPU was
typically faster only by a factor of 2-10.

The relative performance of GPUs tends to improve with increasing
problem sizes, with increasing number of cells or, as in
Fig.~\ref{fig:PS_stat}, with increasing number of photon packages.
However, there were also exceptions. In the case of point source
simulations (Fig.~\ref{fig:PS_stat}c), this could be caused by the
increasing overheads in GPU atomic operations when many threads are
simultaneously updating a small number of cells close to the source.
The same explanation may also apply to the problems seen when using
emission weighting on GPUs. Although emission weighting results in
lower noise for a given \nP, on GPU it increased (at \nP$\sim 10^6$)
the run times by up to a factor of a few (Fig.~\ref{fig:3Dstat}c).
Also in this case, most updates concern a small number of cells, those
with the highest dust temperatures. The effect is accentuated by high
optical depths that result in frequent scatterings and thus even more
frequent updates. These effects will be smaller for models with
multiple point sources, larger number of cells, and lower optical
depths. The problem could be alleviated by interleaving the creation
and tracking of photon packages so that all threads would not create
new photon packages in the same cells and at exactly the same time.
New OpenCL versions with native support for atomic operations should
further decrease the overheads. In Fig.~\ref{fig:3Dstat}c, the problem
disappears for large \nP\, but this is only a side effect from the
limit on the maximum number of photon packages emitted from any single
cell. 

In contrast with the GPU results, on CPU the calculations were faster
when emission weighting was used. This could result from better cache
utilisation, similar to the factor of $\sim$2 effect observed in
\citet{Lunttila2012}. In OpenCL, the threads belonging to the same
work group perform computations in lockstep. All threads thus create
and perform initial tracking of photon packages at the same time and
within a small volume around the embedded radiation source. With a
smaller number of threads and larger cache memories, the net effect 
can be positive on CPUs. The data for the most frequently accessed
cells may remain in cache during a whole run, but it is difficult to
say if this alone explains the observed factor of $\sim$5 speed up
(Fig.~\ref{fig:3Dstat}c). The above examples at least demonstrate
that, although the same program can be run on both CPUs and GPUs, best
performance may require different algorithm optimisations on different
platforms. 

%% Further improvements should be possible, especially on GPUs where a
%% large number of threads (e.g. 32 or 64) are run in lock-step. In SOC,
%% each photon package is computed by a single thread. As the photon
%% packages of individual threads exit the model volume (or are otherwise
%% terminated), threads wait idle until even the last one of that work
%% group (or ``warp'') has completed its calculations. This can be
%% avoided by interleaving the creation of new photon packages (when
%% needed) and the basic photon package tracking within a single
%% simulation loop. The idea was briefly tested but but no improvements
%% were observed. Nevertheless, the large gap between the theoretical and
%% actual performance suggests that some improvements are possible.
%%
%% In particular, tests showed that the current implementation of
%% weighted sampling for dust re-emission does not work well on GPUs.
%% Another test where the GPU performance was relatively poor (although
%% still better than that of CPU) was the simulation of point source
%% emission. This may yet be improved by changes in the simulation scheme
%% or in the underlying OpenCL libraries, via reduced overheads of atomic
%% operations. 

\section{Conclusions}  \label{sect:conclusions}

The ability of SOC to produce correct results in dust radiative
transfer problems was already tested in \citet{Gordon2017_TRUST-I}. In
this paper, we concentrated on the performance of SOC and investigated
methods that could speed up the computations in problems where the
final dust temperatures are obtained only after several iterations.
The tests resulted in the following conclusions:
\begin{itemize}
\item
SOC performs well in comparison to e.g. the CRT program
\citep{Juvela2005} and modern GPUs provide a factor of 2--10 speed-up
over a multicore CPU. Further reduction of run times should be
possible by fine-tuning the algorithms.
\item
The noise of the temperature solution behaves as expected: it is
inversely proportional to the square root of photon packages and
increases by a factor of two for each additional level of the spatial
grid hierarchy. The tested photon splitting scheme did not
significantly reduce the noise of high hierarchy levels relative to
the run time.
\item
In the test cases, ALI provided moderate acceleration for the
convergence of dust temperature values. The run time overhead varied
from case to case but was of the order of 10\%, small compared to the
potential benefits from the faster convergence.
\item
The use of a reference field without ALI was found to be the most
robust alternative. The desired accuracy could be reached in the least
amount of time by doing the initial iterations with fewer photon
packages.
\end{itemize}
SOC will be developed further to accommodate multiple dust
populations and in Appendix~\ref{sect:SHG} we already discuss tests on
the use of GPUs to speed up to the calculations of stochastically
heated grain emission.

\begin{acknowledgements}
We acknowledge the support of the Academy of Finland Grant No. 285769.
\end{acknowledgements}

\bibliography{MJ_bib}

\begin{appendix}

\section{Calculations with stochastically heated grains} \label{sect:SHG}

SOC concentrates on solving the radiative transfer problem and only
solves grain temperatures for grains in equilibrium with the radiation
field. For stochastically heated grains, these computations are
delegated to an external program, such as the one that in
\cite{Camps2015} was used in connection with the CRT code.  We report
here some results from tests on implementing similar routines with
Python and OpenCL. We refer to the old code as the C program and the
new one as the OpenCL program.

Tests were made using a single dust component and a three-dimensional
model cloud that consisted of 8000 cells, sufficient to get accurate
run-time measurements. The dust corresponds to astronomical silicates
as defined in the DustEM package \cite{Compiegne2011}. The grain size
distribution extends from 4\,nm to 2\,$\mu$m is discretised into 25
logarithmic size bins. All grains were treated in the tests as
stochastically heated while in real applications the largest grains
could be handled using the equilibrium temperature approximation. The
calculations employed a logarithmic discretisation of enthalpies that in
temperature extend from 4\,K to 150\,K for the largest and from 4\,K
to 2500\,K for the smallest grains. The model cloud is illuminated by
the standard interstellar radiation field \citep{Mathis1983}.  The
column density of the cloud varies from $N({\rm H}_2) \sim 3 \times
10^{19}\, {\rm cm}^{-2}$ to $N({\rm H}_2)= 1 \times 10^{22}\, {\rm
cm}^{-2}$. This ensures a wide range of radiation field intensities,
especially for the shorter wavelengths that are relevant for
mid-infrared dust emission. Radiative transfer simulations used a
logarithmic grid of 128 frequencies between the Lyman limit and 2\,mm.

We used as the reference the C program and its routine that solves the
dust temperatures under the ``thermal continuous'' cooling
approximation as described in \citet{Draine2001}. The calculation
reduces to a linear set of equations where the unknowns are the
fraction of dust in each enthalpy bin. The C program pre-calculates
weights for the integration of the absorbed energy over frequency as
well as the cooling rates that in the thermal continuous approximation
only extend downwards to the next enthalpy bin. When the dust emission
is solved cell-by-cell, the program gets a vector of radiation field
intensities (in practice number of absorbed photons). In the
transition matrix $R$, the upward transitions ($R_{j ,i}$, $i<j$) are
obtained by taking a vector product of the intensity with the
integration weights. The pre-calculated cooling rates occupy the
elements $R_{i-1,i}$ above the main diagonal. The diagonal elements
$R_{i,i}$ are determined from detailed balance and are equal to the
sum of the other column elements multiplied by minus one.  The matrix
elements $R_{j,i}=0$, $i>j$ are zeros, which means that the linear
equations can be solved efficiently with forward-substitutions. The C
program is parallelised with OpenMP, further optimisations including
the use of SSE instructions.

In the OpenCL programme we tested first the use of an iterative
solver. Iterative solvers are attractive because one is typically
dealing with large 3D models where the neighbouring cells are
subjected to a very similar radiation field. By using the solution of
the preceding cell as the starting values for the next one, it is
likely that the number of iterations can be kept small. We tested only
basic Gauss-Seidel iterations with Jacobi (diagonal) preconditioning.
By using the solution of the first cell to start the iterations for
each of the other cells, less than ten iterations and a maximum of
$\sim$50 were needed to reach an accuracy that in the final surface
brightness maps translated to rms errors of a couple of per cent (over
a spectral range spanning more than 10 orders of magnitude in absolute
intensity). On the CPU, using the same number of CPU cores, the run
time was nearly identical to that of the C program. On a GPU, the
OpenCL version was faster by a factor of $\sim$5. In the case of the 
thermal continuous approximation, the explicit solution is already very
fast. Therefore, in a general case ($R_{j,i}\neq 0$ for $j<i$), 
iterative solvers should be a good option.

We could not exactly reproduce the results of the C program even with
more iterations. This may be due to the use single-precision floating
point numbers. The use of double precision would slow down the GPU
computations, depending on the hardware. Furthermore, although
Gauss-Seidel iterations provided a sufficient (but still a
low-precision) solution in just a few iterations, the convergence is
not guaranteed. The rate matrix is not diagonally dominant and the
spectral radius of the iteration matrix was either very close to or
even larger than one. The iterations should thus eventually diverge,
which was indeed observed if continued further beyond one hundred
iterations. However, when the problem is similar for a very large
number of cells, one could calculate better preconditioners with a
small per-cell cost.

We also implemented an explicit forward-substitution algorithm similar
to that of the C programme. This is the natural option in the case of
the thermal continuous approximation. The routine worked reliably but
only when part of the operations were performed in double precision.
The use of double precision resulted in some 25\% increase in the GPU
run times but had no effect on CPUs. The results were identical to the
C program, almost down to the machine precision. Because the 
algorithm is faster, the run times were shorter than for the iterative
solver. Compared to the C programme, the OpenCL routine was 5.6 times
faster when run on CPU and 21 times faster when run on GPU. The
speed-up on the CPU suggests that the parallelisation of the C
programme was not optimal but the results also depend on other
factors, such as the memory and disk access patterns. On the laptop,
the OpenCL program ran $\sim$4 times faster on the GPU than on the
six-core CPU.

The actual wall-clock run time on the laptop GPU was some 0.5\,ms per
cell, which includes the calculations of the dust temperature
distributions and the resulting dust emission at 128 frequencies.
Assuming that modelling involved four dust populations, this would
translate to a rate of 500 cells per second and a run time of some
half an hour for a model of $10^6$ cells. The time required for the
radiative transfer simulations is of the same order of magnitude, the
exact balance depending on the chosen frequency, enthalpy, and grain
size discretisations. The costs of radiative transfer and dust
temperature calculations both also scale approximately linearly with
the number of cells. It is therefore already feasible to directly
solve the emission of stochastically heated grains even for relatively
large 3D models. Table look-up methods are still relevant because they
also reduce number of frequencies at which the radiation field needs
to be estimated \citep{Juvela2003, Baes2011}. Nevertheless, GPUs can
speed up the construction of large look-up tables and thus improve the
accuracy of also those methods.

\section{Examples of surface brightness maps} \label{sect:maps}

In Sect.~\ref{sect:results} we examined the accuracy of SOC results
mainly in terms of dust temperature. Here we present examples of
surface brightness maps that correspond to the tests in
Figs.~\ref{fig:T_vs_iter} and ~\ref{fig:plot_RF}, the final result
after 20 iterations. Surface brightness errors are shown by plotting
the difference of two independent runs.

\begin{figure}
%% \sidecaption
\includegraphics[width=8.8cm]{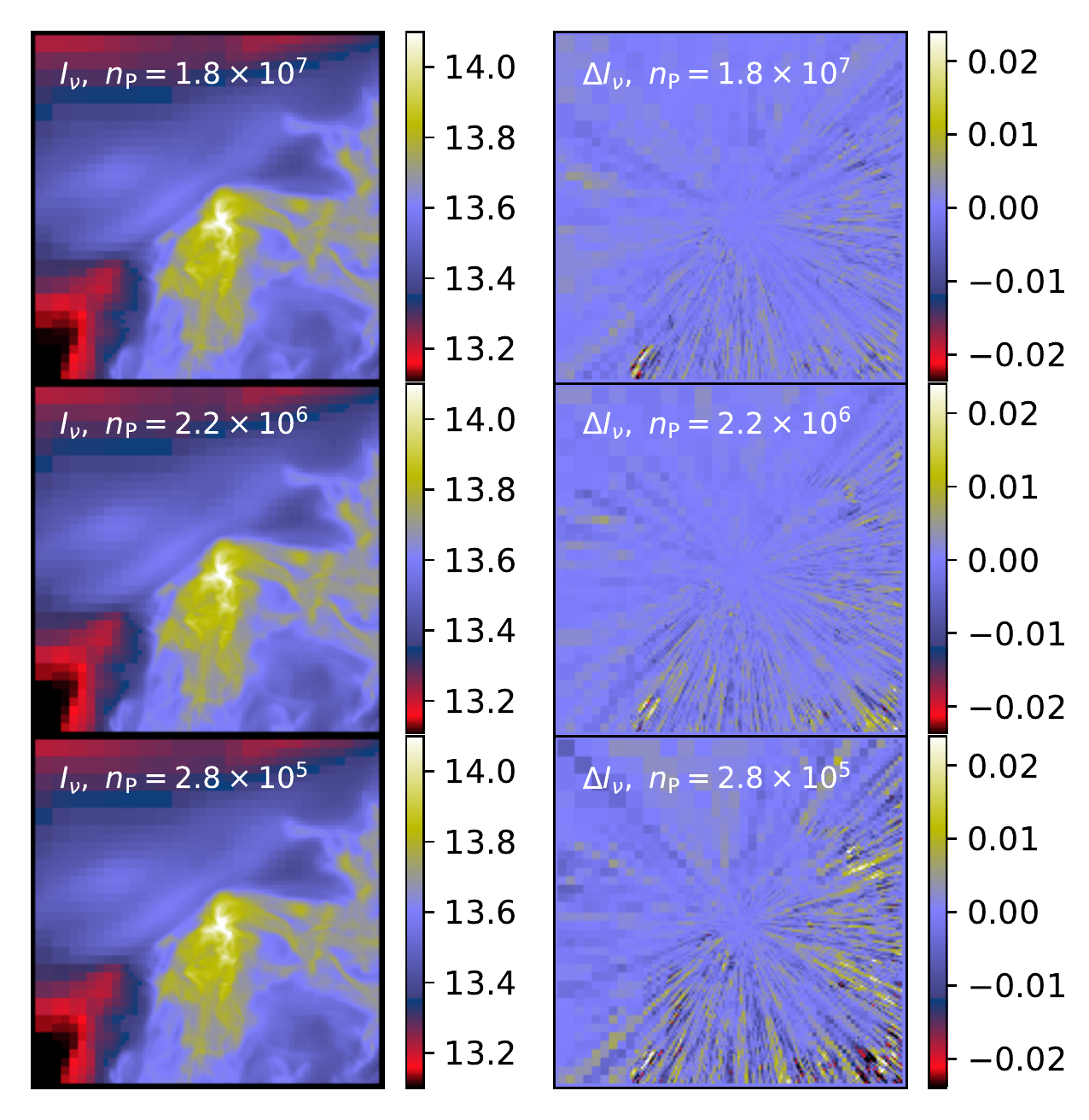}
\caption{
Surface brightness maps at 100\,$\mu$m for the final iteration of
the non-ALI runs of Fig.~\ref{fig:T_vs_iter}. The three rows
correspond, respectively, to the default \nP\, value of $1.8\times
10^7$ and to \nP\, values smaller by a factor of 8 or 64. The left
hand frames show the 100\,$\mu$m surface brightness
$I_{\nu}(100\,\mu{\rm m})$ and the right hand frames the difference
$\Delta I_{\nu}(100\,\mu{\rm m})$ between two identical runs. All maps
are in units of $10^8$\,Jy\,sr$^{-1}$.
}
\label{fig:maps1}
\end{figure}

Figure~\ref{fig:maps1} shows the maps that correspond to
Fig.~\ref{fig:T_vs_iter}, the runs without ALI where the number of
photon packages per iteration was 18 million or smaller by a factor of
8 or 64. The noise is small enough to be visible only in the
difference maps that here only contain errors from the simulation of
the re-radiated dust emission. The error maps show a radial pattern
because hot dust is found only close to the central point source. The
relative contribution of re-radiated dust emission is smaller in
directions where more of the point source radiation escapes the
central clump. This explains the general asymmetry of the pattern
where the errors tend to be smaller on the upper left hand side.

\begin{figure}
%% \sidecaption
\includegraphics[width=8.8cm]{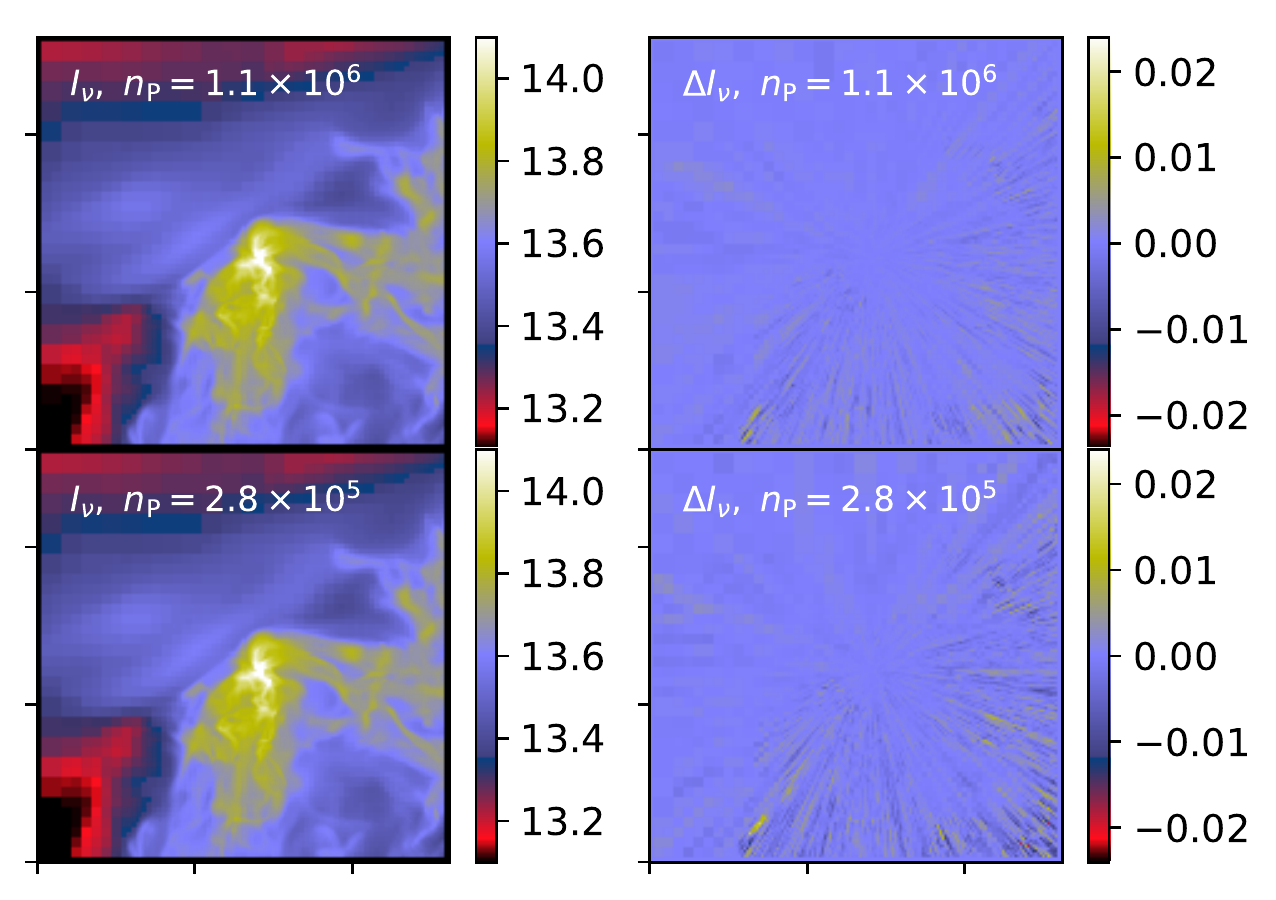}
\caption{
As Fig.~\ref{fig:maps1} but for the runs of Fig.~\ref{fig:plot_RF}
with a reference field. The number of photon packages is
\nP=$1.1\times 10^6$ (upper frames) or four time smaller (lower frames). Maps are
in units of $10^8$\,Jy\,sr$^{-1}$.
}
\label{fig:maps2}
\end{figure}

Figure~\ref{fig:maps2} shows the results when also the reference field
technique is used. The upper frames correspond directly to the runs in
Fig.~\ref{fig:plot_RF}.

\end{appendix}

\end{document}